\documentclass[preprint,notoc]{JHEP3}
\usepackage{graphicx}
\usepackage{amsmath}
\usepackage{psfrag}
\usepackage[sort&compress,numbers]{natbib}

\def\e{\epsilon}
\def \muint{\int \frac{d^{-2\epsilon}\mu}{(2\pi)^{-2\epsilon}}\int \frac{d^4 \ell}{(2\pi)^4}}
\def\spa#1.#2{\left\langle#1\,#2\right\rangle}
\def\spb#1.#2{\left[#1\,#2\right]}
\def\spab#1.#2.#3{\left\langle#1\,#2\,#3\right]}
\def\lo{\ell_1}
\def\lt{\ell_2}

\def\spone{\spab 1.|\ell|.\eta}
\def\sptwo{\spab 2.|\ell_1|.\eta}
\def\spthree{\spab 3.|\ell_2|.\eta}
\def\spfour{\spab 4.|\ell_3|.\eta}
\def\ip{(i+1)}
\def\jp{(j+1)}

\def\spa#1.#2{\left\langle#1\,#2\right\rangle}
\def\spb#1.#2{\left[#1\,#2\right]}
\def\spab#1.#2.#3{\left\langle#1\,#2\,#3\right]}
\def\spba#1.#2.#3{\left[#1\,#2\,#3\right\rangle}
\def\spaa#1.#2.#3.#4.#5.#6{\left\langle#1\,#2\,#3\,#4\.#5\,#6\right\rangle}
\def\spbb#1.#2.#3.#4.#5.#6{\left [#1\,#2\,#3\,#4\,#5\,#6\right ]}
\def\spbnin#1.#2.#3.#4.#5.#6.#7.#8.#9{\left [#1|\,#2\,#3\,#4\,#5\,#6\,#7\,#8|\,#9\right \rangle}
\def\spanin#1.#2.#3.#4.#5.#6.#7.#8.#9{\left \langle#1|\,#2\,#3\,#4\,#5\,#6\,#7\,#8|\,#9\right]}
\def\spasev#1.#2.#3.#4.#5.#6.#7{\left \langle#1|\,#2\,#3\,#4\,#5\,#6\,|#7\right]}
\def\spafiv#1.#2.#3.#4.#5{\left \langle#1|\,#2\,#3\,#4\,|#5\right]}
\def\spbsev#1.#2.#3.#4.#5.#6.#7{\left [#1|\,#2\,#3\,#4\,#5\,#6|\,#7\right \rangle}
\def\spbsix#1.#2.#3.#4.#5.#6{\left [#1|\,#2\,#3\,#4\,#5|\,#6\right ]}
\def\spbfiv#1.#2.#3.#4.#5{\left [#1|\,#2\,#3\,#4|\,#5\right \rangle}
\def\spbf#1.#2.#3.#4{\left [#1\,#2\,#3\,#4\right ]}
\def\spahr#1.#2{\langle#1\,\hat{#2}\rangle}
\def\spaah#1.#2.#3.#4{\langle#1\,#2\,#3\,\hat{#4}\rangle}
\def\spaahl#1.#2.#3.#4{\langle\hat{#1}\,#2\,#3\,#4\rangle}
\def\spabh#1.#2.#3{\langle#1\,\widehat{#2}\,#3]}
\def\spahl#1.#2{\langle\hat{#1}\,#2\rangle}
\def\spahh#1.#2{\langle\hat{#1}\,\hat{#2}\rangle}
\def\spaas#1.#2.#3{\left\langle#1\,#2\,#3\right\rangle}
\def\spbhl#1.#2{\left[\hat{#1}\,#2\right]}
\def\spbhr#1.#2{\left[#1\,\hat{#2}\right]}
\def\spabhh#1.#2.#3{\langle\hat{#1}\,\hat{#2}\,#3]}
\def\spbah#1.#2.#3{[#1\,\widehat{#2}\,#3\rangle}
\def\n{\eta}
\def\lo{\ell_1}

\def\lt{\ell_2}

\def\ip{(i+1)}
\def\im{(i-1)}
\def\jp{(j+1)}

\def\nm{(n-1)}

\def\NN{$\mathcal{N}=4$}
\DeclareMathOperator{\F}{\mathit{F}}

\DeclareMathOperator{\tr}{ {\rm tr}}
\def\trm{\tr_-}
\def\trp{\tr_+}

\DeclareMathOperator{\Ftme}{ {\rm F}^{2me}_4}

\DeclareMathOperator{\Fom}{ {\rm F}^{1m}_4}
\def\FF#1{\sideset{_2}{_1}\F\bigg(#1\bigg)}

\def\mc#1{\mathcal{#1}}
\def\sl#1{\slash\!\!\!{#1}}
\def\hl#1.#2{\langle\hat{#1}#2\rangle}
\def\hr#1.#2{\langle#1\hat{#2}\rangle}
\def\bhl#1.#2.#3.#4{\langle #1\widehat{#2}#3\hat{#4}\rangle}
\def\thl#1.#2.#3{\langle#1\widehat{#2}#3]}
\def\spabt#1.#2.#3{$\begin{tiny}$\langle#1\,#2\,#3]$\end{tiny}$}

\def\intOm#1.#2{\int \frac{d\Omega#1}{#2}}

\preprint{
  IPPP/08/66\\
  \\
  \today}

\title{One-Loop Gluonic Amplitudes from Single Unitarity Cuts}

\author{E. W. Nigel Glover,
    \ Ciaran Williams \ 
    	\\
	Department of Physics, University of Durham, Durham, DH1 3LE, UK
	\\
	E-mails: 
        {\tt e.w.n.glover@durham.ac.uk}, 
        	{\tt ciaran.williams@durham.ac.uk}.
}

\abstract{
We show that one-loop amplitudes in massless gauge theories can be determined from single cuts. 
By cutting a single propagator and putting it on-shell, the integrand of 
an $n$-point one-loop integral is transformed into an $(n+2)$-particle 
tree level amplitude.   
The single-cut approach described here is complementary to the double or multiple unitarity cut approaches commonly used in the literature. 
In common with these approaches, if the cut is taken in four dimensions,  
one finds only the cut-constructible parts of the amplitude, while if the cut is in $D=4-2\epsilon$ dimensions, both rational and cut-constructible parts are obtained.   
We test our method by reproducing the known results for the fully rational all-plus and mostly-plus QCD amplitudes, $A^{(1)}_4(1^+,2^+,3^+,4^+)$ and $A^{(1)}_5(1^+,2^+,3^+,4^+,5^+)$.
We also rederive expressions for the scalar loop contribution to the
four-gluon MHV amplitude, $A_4^{(1,\mc{N}=0)}(-,-,+,+)$ which has both
cut-constructible and rational contributions, and the fully cut-constructible
$n$-gluon MHV amplitude in $\mc{N}=4$ Supersymetric Yang-Mills, $A_4^{(1,\mc{N}=4
)}(-,-,+,\ldots,+)$.
\\
\today
}


\begin{document}

\section{Introduction}

As the first collisions at the Large Hadron Collider (LHC) draw nearer, there is a growing sense of optimism in the particle physics community that a new era of discovery and exploration awaits. 
In order to make sense of the vast amounts of data expected from the LHC, precise knowledge is required regarding the Standard  Model processes which form a background to all new physics signals. For many QCD and electroweak events involving multi-particle final states, tree-level calculations fail to generate results of sufficient accuracy and next to leading order (NLO) corrections are required \cite{Bern:2008ef}. Over the last few years it has become apparent that unitarity-based methods offer many advantages over more traditional approaches in the calculation of one-loop amplitudes in massless theories. In particular, unitarity methods employ on-shell quantities as the fundamental building blocks which can result in considerable simplification of the intermediate stages of the calculation. 
\\

Inspired by Witten's discovery \cite{Witten:twstr} of the simplicity of maximally helicity violating (MHV) amplitudes in Penrose's twistor space, Cachazo, Svr\v{c}ek and Witten \cite{Cachazo:2004by} developed a novel diagrammatic technique for construction of tree level amplitudes in Yang-Mills gauge theory. By taking the Parke-Taylor \cite{Parke:ngluon} (MHV) amplitudes, 
\begin{eqnarray}
A_n(1^+,\dots i^-,\dots,j^-,\dots,n^+)=i\frac{\spa i.j^4}{\spa1.2\spa2.3\dots\spa n.1},
\end{eqnarray}
off-shell the authors were able to use MHV amplitudes as vertices to construct amplitudes containing increasing numbers of negative helicity gluons.  The off-shell vertices are connected by scalar propagators which link gluons of opposite helicity. These methods have been successfully extended to generate amplitudes for a wide range of massless theories \cite{Wu:2004fba,Wu:2004jxa,Georgiou:2004by,Georgiou:2004wu}, amplitudes involving  Higgs bosons \cite{Dixon:MHVhiggs,Badger:MHVhiggs2} and massive vector boson currents \cite{Bern:EWcurrents}. Recently the CSW prescription has been extended to include the generation of tree amplitudes containing massive (coloured) scalars  \cite{Boels:2007pj,Boels:2008ef} and fermions \cite{Schwinn:2008fm}. Progress at one-loop was stimulated by the work of Brandhuber, Spence and Travaglini \cite{Brandhuber:n4}, who observed that at one-loop, one can relate the MHV rules to a dispersive integral over phase space. This lead to the calculation of several one-loop $n$-point MHV amplitudes, both in supersymmetric \cite{Brandhuber:n4,Bedford:n1,Quigley:2004pw,Glover:2008tu}, and non-supersymmetric theories \cite{Bedford:nonsusy,Badger:2007si,Glover:2008ffa}. 
\\ 

Unitarity in its modern form owes its origins to the work of Bern, Dixon, Dunbar and Kosower \cite{BDDK:uni2,BDDK:uni1} who in the mid nineties used the simplicity of tree-level amplitudes in Yang-Mills to correctly reconstruct the discontinuity structure of many one-loop amplitudes in $\mc{N}=4$ and $\mc{N}=1$. The original work required the cut propagators and the tree amplitudes to be four-dimensional, and as a result  QCD amplitudes were unable to be fully reconstructed. The parts of the amplitude found by four-dimensional cuts were called cut-constructible, whilst the elusive remaining terms were deemed rational (since they had no discontinuities in physical invariants).  
\\

The search for a method which calculated amplitudes fully from on-shell methods has led to the development of many new ideas and techniques. The discovery of novel recursion relations in gauge theories by Britto, Cachazo and Feng  \cite{Britto:rec} (and later proven with Witten~\cite{Britto:proof}) inspired the development of the unitarity bootstrap \cite{Bern:bootstrap,Berger:genhels}. This method relied on the fact that the rational terms, having no logarithms or branch cuts, obeyed similar recursive relations to tree-level amplitudes. This meant that the cut-constructible terms were calculated from four-dimensional unitarity and rational terms could be calculated using recursion relations. A major advantage of the bootstrap is that rational and cut-constructible terms are both calculated in four dimensions, which allows the use of the spinor helicity formalism and the simplifcations associated with it. 
Recently, the bootstrap method has been implemented in the program {\tt Blackhat} \cite{Berger:2008sj}
which calculates NLO amplitudes numerically. 
\\

It was noted long ago \cite{Bern:1995db} that if one performed the unitarity cut in $D$ dimensions the amplitude could be fully reconstructed. Over the last few years the development of efficient
four-dimensional unitarity based methods, such as generalised unitarity and spinor integration \cite{Britto:genuni,Britto:sqcd,Britto:ccqcd}, have inspired the development of new $D$-dimensional unitarity methods. Extensions of the multiple cut approach of generalised unitarity to $D$ dimensions \cite{Brandhuber:2005jw} were able to correctly reproduce known QCD rational amplitudes. In a series of papers, Britto, Feng and collaborators \cite{Anastasiou:DuniII,Anastasiou:DuniI,Britto:Duni,Britto:2008sw,Britto:2008vq,Feng:2008ju} implemented spinor integration in $D$ dimensions (and also extended the applications to include massive theories). Recently \cite{Badger:2008cm} a implementation of $D$-dimensional unitarity has been proposed which should allow direct calculation of the rational pieces of amplitudes. \\

Inspired by the developments in analytic unitarity based calculations,  numerical tools for evaluating loop-amplitudes have made enormous progress.  In a series of remarkable papers Ossola, Pittau and Papadopoulos developed an algebraic method  \cite{Ossola:2006us,Ossola:2008xq,Mastrolia:2008jb} for extracting the coefficients of the loop-integrals at the integrand level, together with the rational parts. In the last year this has led to the release of {\tt CutTools} \cite{Ossola:2007ax} a program which numerically calculates loop amplitudes and has been used to study six-photon amplitudes~\cite{Ossola:2007bb} as well as the NLO QCD corrections to tri-vector boson production \cite{Binoth:2008kt}. Another program which is based upon $D$-dimensional unitarity is {\tt Rocket}~\cite{Giele:2008bc}, which arose from the observations of \cite{Giele:2008ve,Ellis:2007br} where it has been shown that one can write a generic loop amplitude as an expansion of $D$, $D+2$ and $D+4$ master integrals. A recent achievement has been the calculation of scattering amplitudes involving up to 20 gluons~\cite{Giele:2008bc}. The method has also been extended to incorporate massive fermions \cite{Ellis:2008ir}
and vector bosons \cite{Ellis:2008vecbos}.\\

When undertaking  calculations involving gluon loops, one can make use of the following supersymmetric breakdown,
\begin{equation}
\mc{A}_g=\mc{A}^{\mc{N}=4}-4\mc{A}^{\mc{N}=1}+\mc{A}^{\mc{N}=0, scal}.
\end{equation}
The important realisation being that $\mc{N}=4$ and $\mc{N}=1$ theories are cut-constructible 
 so that $D$-dimensional unitarity is only required to calculate the last term in the above equation. If the four-dimensional helicity scheme (FDH) is used, external momenta can be kept in four dimensions, whilst the internal loop momenta become $D$-dimensional. In order to preserve the techniques useful in four dimensions it is common to separate the loop momentum into 4 and $-2\epsilon$ components. Since the external momenta are four-dimensional the $-2\epsilon$ dimensions decouple and behave like a mass, which must be integrated over. This transformation makes it very useful to know on-shell amplitudes involving gluons and massive (coloured) scalars. These amplitudes have been derived using the recursion relations \cite{Badger:massrec, Forde:masssclr} and more recently they were derived in the CSW formalism \cite{Boels:2007pj,Boels:2008ef}.
\\

The concept of cutting a single propagator in loop amplitudes is not new. However until now, most studies of single cuts have been linked with the 
Feynman tree theorem \cite{Brandhuber:2005kd}. This theorem states that a generic one-loop amplitude can be written as a sum over its allowed cuts, from single cuts upwards. Recently it was proposed \cite{Catani:2008xa} that entire amplitudes could be calculated at one-loop from single cuts avoiding the FTT by a using a specially deformed contour of integration in the complex plane. The method we use in this paper is directed more along the lines of generalised unitarity. We apply the cut to both the one-loop integral and the one-loop basis functions and use reduction techniques to extract the coefficients of the master integrals appearing in the one-loop expansion. At the final stages of the calculation we can simply re-insert the cut propagator and return to the basis of scalar integrals. 
\\

This paper proceeds as follows, section 2 introduces the single cut method we will employ and reviews the integral basis used in $D$-dimensional unitarity. Section 3 briefly reviews the construction of tree amplitudes using the CSW rules with massive scalars derived by Boels and Schwinn.  Sections 4 and 5 contain detailed example calculations. We calculate the four- and five-gluon all plus amplitudes in section 4. In section 5 we extend the examples to include four-gluon amplitudes with one or two gluons of negative helicity. We also give an example of a fully cut-constructible amplitude by calculating the $n$-gluon MHV amplitude with adjacent negative helicity gluons in $\mc{N}=4$ SYM. In Section 6 we draw our conclusions. For completeness, we list the $\e$ dependence of the scalar integrals we encounter as well as the notation we use throughout the paper in appendix A. 

\section{Single cuts and unitarity}

\subsection{Single cuts} 

The main result of this paper is that colour-ordered multi-gluon one-loop amplitudes can be calculated from single cuts. In a similar manner to previous unitarity studies we find that if four-dimensional tree amplitudes are used we can reconstruct only the cut-constructible pieces of an amplitude. However, if the cut is performed in  $D=4-2\epsilon$ dimensions then it is possible to fully determine the amplitude. In this section we describe how the single cut works and how to implement this concept within $D$-dimensional unitarity. 

We begin by considering an arbitrary one-loop processes in $D$ dimensions with massless propagators which can be written in the following manner,
\begin{eqnarray}
A^{(1-loop),D}_{n}=\sum_{j=2}^{n} c^{D}_i(\{p_k\})I^{D,i}_j(\{p_k\}).
\label{eq:bas}
\end{eqnarray}
Here $c^{(D)}$ are coefficients which depend only on the dimension $D$ of the loop momentum  and a set of outgoing external momenta $\{p_k\}$ which we take to be four-dimensional. The scalar $n$-point integral in $D$ dimensions is defined by 
\begin{eqnarray}
I_n^{D}[1]= i(-1)^{n+1}(4\pi)^{D/2}\int \frac{d^{D} L}{(2\pi)^{D}} \frac{1}{(L^2+i0)((L-p_1)^2+i0)\dots((L-\sum_{i=1}^{n-1}p_i)^2+i0)},\nonumber \\
\label{eqn:scain}
\end{eqnarray}
where the $+i0$ denotes the prescription for continuing the pole off the real axis.  From now on, we will not show the $+i0$ explicitly. 
Finally, the implicit summation over $i$ represents the range of allowed momentum configurations in the denominator for a $j$-point scalar integral (i.e. for a four-point we sum over four, three, two and one mass boxes). In massless theories, scalar tadpoles $(I_1)$ and on-shell bubbles $(I_2(p^2_i))$ vanish regardless of $D$. 

\begin{figure}
\begin{center}
\psfrag{a}{$-L$}
\psfrag{b}{$L$}
\psfrag{i}{$i$}
\psfrag{im}{$i-1$}
\includegraphics[height=4cm]{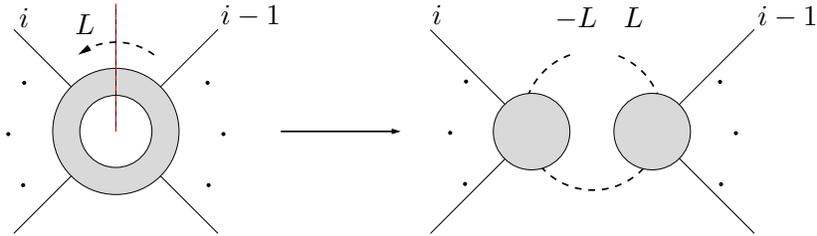}
\caption{Schematic of a the $C_{i,i-1}$ single cut.   The propagator between momenta $p_i$ and $p_{i-1}$ is cut, leading to two additional (dashed) outgoing external scalar particles. This cut is not sensitive to scalar integrals that have $p_i$ and $p_{i-1}$ emitted from the same vertex.}
\label{fig:cut}
\end{center}
\end{figure}

We now consider what happens to both sides of \eqref{eq:bas} when we take a single cut of the loop integral.  We choose to make the cut between the external momenta $p_i$ and $p_{i-1}$ and we denote this to be the $C_{i,i-1}$ cut as shown in Fig.~\ref{fig:cut}.   We follow the usual procedure of taking the  propagator on-shell via the replacement
\begin{eqnarray}
\frac{i}{L^2} \rightarrow \delta{(L^2)}.
\end{eqnarray}

On the left-hand side of \eqref{eq:bas} the integrand becomes an $(n+2)$-particle tree level amplitude,
so that
\begin{eqnarray}
A^{(1-loop),D}_{n} \rightarrow \int \frac{d^{D} L}{(2\pi)^{D}} \delta(L^2) A^{(0),D}_{(n+2)}(-L,p_i,p_{i+1},\dots,p_{i-1},L)
\label{eq:def}
\end{eqnarray}
as illustrated in Fig.~\ref{fig:cut}.
We note that the sum of the external momenta $\sum_{j=i}^{i-1} p_j$ vanishes, just as for the loop amplitude.
This is the definition of a single cut in the $C_{i,i-1}$ channel. The integrand is an $(n+2)$-particle tree amplitude which is manifestly gauge invariant. This is a very useful feature of the single cut since a different gauge choice can be made for each cut.

\begin{figure}
\begin{center}
\psfrag{A}{$(a)$}
\psfrag{B}{$(b)$}
\psfrag{i}{$i$}
\psfrag{im}{$i-1$}
\includegraphics[height=4cm]{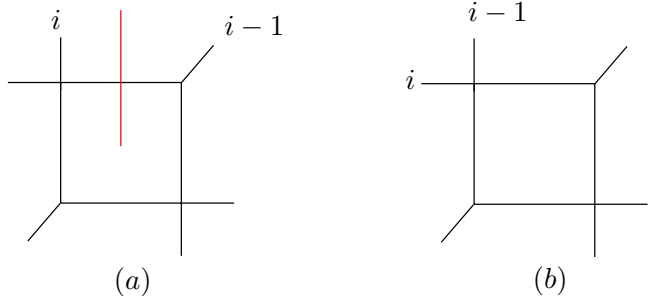}
\caption{Examples of master integrals which  $(a)$  can and  $(b)$ cannot be found by a $C_{i,i-1}$ cut.}
\label{fig:box1}
\end{center}
\end{figure}

A single cut simultaneously determines the coefficients of many of the scalar basis integrals, since every basis integral has some form of single cut. It is logical to question which coefficients can be determined from a given cut. The important quantity is the ordered set of momenta, since this determines which propagator has been cut.  For example in the $C_{i,i-1}$ cut only 
scalar integrals which have $p_i$ and $p_{i-1}$ emitted from different vertices
give a contribution.\footnote{For massive theories, this is not the case as there may be tadpole contributions.}
An example of a basis integral which has a contribution in the $C_{i,i-1}$ cut
is shown in Fig.~\ref{fig:box1}$(a)$. 
In general determination of every coefficient of the basis integrals will always require fewer than $n$ cuts. 

To summarise, to calculate a generic colour ordered $n$-gluon one-loop amplitude from a single cut in the $C_{i,i-1}$ channel, one must first draw all allowed $(n+2)$-particle diagrams where the two new cut particles are placed between $i$ and $i-1$. 

\subsection{Four-dimensional unitarity}

The discussion in the previous section is not sensitive to the specific value of $D$. One can always write a one-loop amplitude as an summation of scalar integrals, multiplied by coefficients which are rational functions of the external momenta and the dimension $D$.
Of course, when one wants to perform an actual calculation the value of $D$ is crucial, since it sets the dimension of the tree inputs. Ideally one would use four-dimensional tree amplitudes, taking advantage of the simplifications associated with the spinor helicity formalism and on-shell techniques, i.e. one would write~\eqref{eq:def} as,
\begin{eqnarray}
A^{(1-loop),D}_{n} \rightarrow \int \frac{d^{4-2\epsilon} L}{(2\pi)^{4-2\epsilon}} \delta(L^2) A^{(0),4}_{(n+2)}(-L,p_i,p_{i+1},\dots,p_{i-1},L).
\end{eqnarray}
Of course, one does not expect to reconstruct the entire integral using four-dimensional trees. Bern, Dixon, Dunbar and Kosower showed \cite{BDDK:uni1,BDDK:uni2} that by using four-dimensional tree inputs when calculating amplitudes with double cuts one misses terms $\mc{O}(\epsilon^0)$. These rational polynomials arise from cancellations of the form $\epsilon \times (1/\epsilon)$ which occur when a $D$-dimensional contraction from tensors in the numerator cancels a divergence which arises from UV structure of the integrand\footnote{UV and IR divergences can be regulated by taking the denominators back into $D=4-2\epsilon$ dimensions.}. However, they were able to show that in $\mc{N}=4$ and $\mc{N}=1$ Supersymmetric Yang-Mills (SYM) theories, these missing terms are uniquely associated with the one-loop basis functions, so that the entire amplitudes could be calculated from four-dimensional trees. For QCD or scalar loop amplitudes however, there rational polynomials exist which can never be determined from four-dimensional unitarity. To fully determine the amplitude, therefore, the tree input must remain in $D$ dimensions.  

\subsection{$D$-dimensional unitarity}

The $D$-dimensional scalar $n$-point function in massless theories is defined in \eqref{eqn:scain} 
where $L$ exists fully in $D=4-2\epsilon$ dimensions and the external sums of momenta are four-dimensional. We can split $L$ into its four ($\ell$) and $-2\epsilon$ ($\mu$) dimensional components, 
\begin{equation}
L=\ell+\mu.
\end{equation}
This transforms the measure as, 
\begin{equation}
\int d^{D} L=\int d^{-2\epsilon}\mu\int d^4\ell,
\end{equation}
while the propagators transform as 
\begin{equation}
(L-p_i)^2 \longrightarrow (\ell-p_1)^2-\mu^2,
\end{equation}
so that the scalar integral becomes
\begin{eqnarray} 
I_n^{D}[1]=\int \frac{d^{-2\epsilon}\mu}{(2\pi)^{-2\epsilon}}\int \frac{d^{4} \ell}{(2\pi)^{4}} \frac{ i(-1)^{n+1}(4\pi)^{2-\epsilon}}{(\ell^2-\mu^2)((\ell-p_1)^2-\mu^2)\dots((\ell-\sum_{i=1}^{n-1}p_i)^2-\mu^2)}.
\end{eqnarray}
In other words, a massless integral in $D$ dimensions has been transformed into a four-dimensional integral where each propagator looks like the propagator of a massive scalar particle.  In the single cut approach we are advocating here, a massive scalar propagator is placed on-shell yielding two external massive scalar particles.    

Integrals with dot products involving $L$ in the numerator can be treated in much the same way. A generic numerator now acquires a polynomial dependence on $\mu^2$. We can systematically use the results of \cite{Bern:1995db} to relate integrals with additional factors of $\mu^2$ in the numerator to higher-dimensional scalar integrals
\begin{eqnarray}
I_n^{D}[(\mu^2)^r]&=&\int \frac{d^{-2\epsilon}\mu}{(2\pi)^{-2\epsilon}}\int \frac{d^{4} \ell}{(2\pi)^{4}} \frac{ i(-1)^{n+1}(4\pi)^{2-\epsilon}(\mu^{2r})}{(\ell^2-\mu^2)((\ell-p_1)^2-\mu^2)\dots((\ell-\sum_{i=1}^{n-1}p_i)^2-\mu^2)}.\nonumber\\
&=&-\epsilon(1-\epsilon)\dots(r-1-\epsilon)I_n^{D+2r}[1].
\end{eqnarray}

Over the past few years many authors have investigated $D$-dimensional unitarity using differing numbers of cuts to determine amplitudes. Several approaches have been shown to correctly reproduce fully the rational parts of QCD amplitudes. These include $D$-dimensional generalised unitarity~\cite{Brandhuber:2005jw} where quadruple and triple cuts were used to determine amplitudes. A $D$-dimensional version of the the triple cut was derived in \cite{Mastrolia:2006ki}. Whilst double cuts using spinor integration in $D$-dimensions have been studied extensively \cite{Anastasiou:DuniII,Anastasiou:DuniI,Britto:Duni,Britto:2008sw,Britto:2008vq}. Recently the $D$-dimensional double cut was used to determine the five-point $A_5^{(1)}(-,+,+,+,+)$ amplitude to  $\mc{O}(\epsilon)$ for the first time \cite{Feng:2008ju}. A common theme in these approaches has been separation of $L$ into 4 and $(D-4)$-dimensional components as discussed above. 

\section{CSW construction with massive scalars}

The single cut technique uses $(n+2)$-point tree amplitudes to calculate $n$-gluon one-loop amplitudes.   Often the two additional particles are massive scalars and therefore we require an efficient method to generate 
tree level amplitudes involving gluons and massive scalars. 
To this end, we use the CSW rules for massive scalars that were derived by Boels and Schwinn \cite{Boels:2007pj,Boels:2008ef} and which we briefly review in this section. 

Using Lagrangian and twistor space methods, the authors of \cite{Boels:2007pj,Boels:2008ef} were able to show that amplitudes involving massive scalars could be generated using the usual massless scalar-gluon vertices,
\begin{eqnarray}
V_{CSW}(1^+,\dots,i^-,\dots,j^-,\dots,n^+)= i \frac{\spa i.j^4}{\spa1.2\spa2.3\dots\spa \nm.n\spa n.1},\\
V_{CSW}(1_{\phi},\dots,i^-,\dots,n_{\phi})=-i\frac{\spa 1.i^2\spa n.i^2}{\spa1.2\spa2.3\dots\spa \nm.n\spa n.1},\\
V_{CSW}(1_{\phi},\dots,i_{\phi},j_{\phi},\dots,n_{\phi})=i\frac{\spa 1.i\spa j.n\spa1.n\spa i.j}{\spa1.2\spa2.3\dots\spa \nm.n\spa n.1}.
\end{eqnarray}
Here $\dots$ represent a cyclic ordering of gluons of positive helicity and $k_{\phi}$ represents a massive scalar with momentum $k$. 
These vertices are supplemented by an additional tower of vertices arising from the scalar mass term in the Lagrangian, which couples any number of positive gluons to a pair of scalars. 
\begin{eqnarray}
V_{CSW}(1_{\phi},2^+,\dots,(n-1)^+,n_{\phi})=-i\mu^2\frac{\spa 1.n}{\spa1.2\dots\spa n-1.n}.
\end{eqnarray}
When scalars emitted at different vertices are connected, 
one uses a massive scalar propagator, 
\begin{eqnarray}
\frac{i}{p^2-\mu^2},
\end{eqnarray}
where $p$ is the momentum carried by the exchanged particle and $\mu$ is the scalar mass. Spinors which are associated with massive scalars or off-shell gluons are continued off-shell via the usual CSW prescription, 
\begin{eqnarray}
\spa i.x\rightarrow \frac{\spab i.|x|.\n}{\spb x.\n}.
\end{eqnarray}
Here $i$ and $\n$ are arbitrary massless spinors subject to the constraint that once $\n$ has been set, it remains fixed in all contributing diagrams. 

\section{One-loop amplitudes of gluons with positive helicity}

\subsection{The all-plus four-gluon amplitude}
\label{sec:allplus4}
The four-point amplitude for gluons with only positive helicity was first calculated using string based methods in \cite{Bern:1l4g}, and it was given to $\mc{O}(\epsilon)$ in  \cite{Bern:1995db}. This helicity amplitude vanishes at tree-level, and is therefore entirely rational.   It is obtained by computing the contribution of a massive complex scalar circulating in the loop. The amplitude is particularly simple \cite{Bern:1995db},  
\begin{eqnarray}
\label{eq:allplus4}
A^{(1)}_4(1^+,2^+,3^+,4^+)=\frac{2i}{(4\pi)^{2-\epsilon}}\frac{\spb 1.2\spb3.4}{\spa1.2\spa3.4} K_4,
\end{eqnarray}
where 
\begin{eqnarray}
K_4 =I_4[\mu^4]=-\epsilon(1-\epsilon)I^{D=8-2\epsilon}_4=-\frac{1}{6}+\mc{O}(\epsilon).
\end{eqnarray}

\begin{figure}
\begin{center}
\psfrag{B}{$(a)$}
\psfrag{C}{$(b)$}
\psfrag{D}{$(e)$}
\psfrag{E}{$(f)$}
\psfrag{A}{$(c)$}
\psfrag{F}{$(d)$}
\psfrag{G}{$(g)$}
\psfrag{a}{${1^+}$}
\psfrag{c}{${2^+}$}
\psfrag{d}{${3^+}$}
\psfrag{b}{${4^+}$}
\includegraphics[width=14cm]{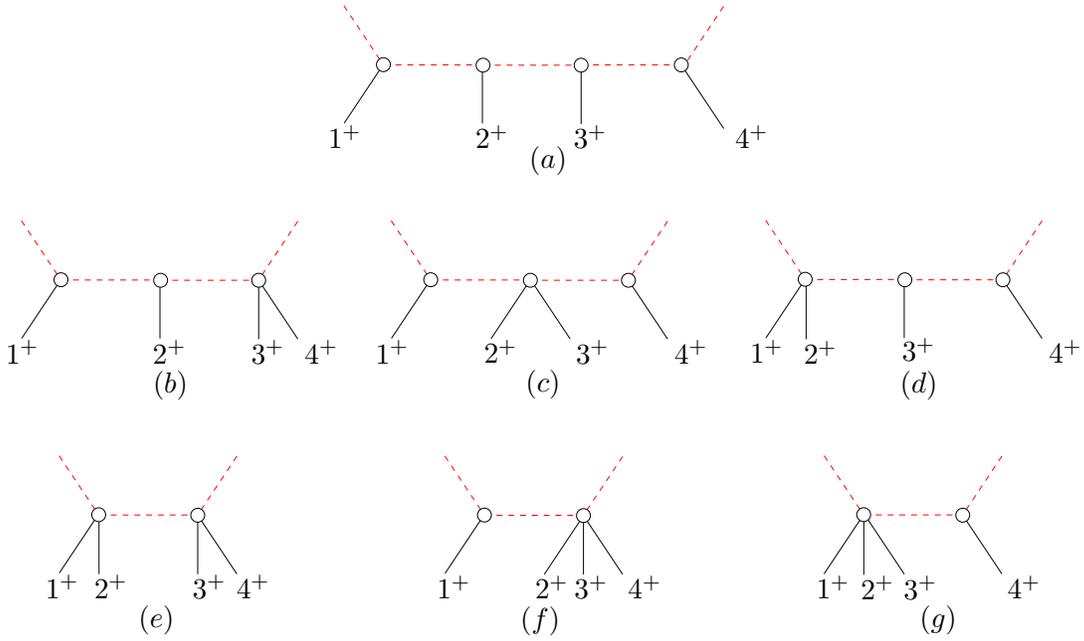}
\caption{The MHV topologies associated with a $C_{1,4}$ cut of the four-gluon all-plus amplitude. Other cuts are obtained from this one by cyclic relabeling of the external particles.}
\label{Fig:plu4}
\end{center}
\end{figure}
The symmetries of the amplitude mean that we can calculate it fully from a solitary single cut, and  we choose to make the $C_{1,4}$ cut.  That is, we put the propagator that connects the external momenta $p_1$ and $p_4$ on-shell. 
Fig.~\ref{Fig:plu4} depicts the four possible diagrams for the four-gluon all-plus amplitude. 

Note that the off-shell prescription for the CSW rules introduces a reference vector $\eta$ which we will always choose such that $|\eta\rangle = |i\rangle$, where $p_i$ is an external momenta.  We can therefore exploit the structure of the three-point $\phi\phi g$ vertex by setting $|\eta\rangle = |4\rangle$ so that diagrams such as Fig.~\ref{Fig:plu4}(a) vanish. We will systematically use this simplification to reduce the number of contributing diagrams.

The MHV rules introduce spurious denominators of the type $\spab \alpha.|\ell_{\alpha-1}|.\n$.  Therefore, to keep track of these denominators (which cannot appear in the final answer),  we introduce the following shorthand notation $i_{\alpha\dots \beta}$ to denote the contribution from a certain diagram $i$ which contains spurious terms $\spab \alpha.|\ell_{\alpha-1}|.\n \dots \spab \beta.|\ell_{\beta-1}|.\n$. The diagram shown Fig.~\ref{Fig:plu4}(b) is therefore labeled as $b_{12}$ since it has spurious poles $\spone$ and $\sptwo$, 
\begin{eqnarray}
b_{12}&=&-\muint \delta(\ell^2-\mu^2) V(-\ell,1^+,\ell_1)\frac{1}{d(\ell_1)}\nonumber\\&&\qquad\qquad\times V(-\ell_1,2^+,\ell_2)\frac{1}{d(\ell_2)}V(-\ell_2,3^+,4^+,\ell)\nonumber\\
&=&-i\intOm[\mu^6].{d(\ell_1)d(\ell_2)} \frac{\spb\n.1\spb\n.2\spb\n.3}{\spa3.4\spone\sptwo \spab 4.|\ell|.\n}\nonumber\\
&=&-i\intOm[\mu^6].{d(\ell_1)d(\ell_2)d(\ell_3)} \frac{\spb\n.1\spb\n.2\spb\n.3}{\spa3.4\spone\sptwo},
\end{eqnarray}
where we follow the notation of Boels and Schwinn~\cite{Boels:2008ef},
\begin{eqnarray}
\ell_i=\ell-\sum_{j=1}^{i}p_j \qquad \mathrm{and} \qquad d(\ell_i)=\ell^2_i-\mu^2 
\end{eqnarray}
and have used the fact that $\spab 4.|\ell|.\n=\spab 4.|\ell|.4=d(\ell_3)$.
The integration measure is given by
\begin{equation}
\int d\Omega[\mu^n]=\muint\delta(\ell^2-\mu^2) \mu^n.
\end{equation}
Diagram Fig.~\ref{Fig:plu4}(e) has the form,
\begin{eqnarray}
e_{12}&=&i\intOm[\mu^4].{d(\ell_2)d(\ell_3)} \frac{\spb\n.3\spbf \n.|\ell.P_{1,2}|.\n}{\spa1.2\spa3.4\spone\sptwo},
\end{eqnarray}
while diagram Fig.~\ref{Fig:plu4}(f) is given by,
\begin{eqnarray}
f_2&=&i\intOm[\mu^4].{d(\ell_1)d(\ell_3)} \frac{\spb\n.1^2}{\spa2.3\spa3.4\sptwo}.
\end{eqnarray}
Combining terms
\begin{eqnarray}
b_{12}+e_{12}=i\intOm[\mu^4].{d(\ell_1)d(\ell_2)d(\ell_3)}\frac{(\mu^2\spbf \n.|1.2|.\n+d(\ell_1)\spbf \n.|\ell.P_{1,2}|.\n)\spb\n.3}{\spa1.2\spa3.4\spone\sptwo} 
\end{eqnarray}
and using the identity \cite{Boels:2008ef} 
\begin{eqnarray}
\mu^2\spbf \n.|i.\ip|.\n&=&\spb i.\ip\spba\n.|\ell_i|.i\spba\n.|\ell_{i+1}|.\ip+d(\ell_{i-1})\spba\n.|\ell_i|.\ip\spb\ip.\n\nonumber\\&&-d(\ell_i)\spbf\n.|\ell_{i-1}.P_{i,i+1}|.\n+d(\ell_{i+1})\spba\n.|\ell_{i-1}|.i\spb i.\n
\label{eq:den}
\end{eqnarray}
with the on-shell condition $d(\ell)=0$, we find that spurious poles of the type $\spone$ are eliminated, 
\begin{eqnarray}
b_{12}+e_{12}=i\intOm[\mu^4].{d(\ell_1)d(\ell_2)d(\ell_3)}\frac{\spb1.2\spb\n.3}{\spa1.2\spa3.4}\nonumber\\
-i\intOm[\mu^4].{d(\ell_1)d(\ell_3)}\frac{\spb\n.1\spb\n.3}{\spa1.2\spa3.4\sptwo}.
\end{eqnarray}
When we combine the above with $f_{2}$ we find that the terms which contain the spurious poles $\sptwo$ also vanish, 
\begin{eqnarray}
b_{12}+e_{12}+f_2=-i\intOm[\mu^4].{d(\ell_1)d(\ell_2)d(\ell_3)}\frac{\spb1.2\spb3.\n}{\spa1.2\spa3.4}.
\label{eq:res4}
\end{eqnarray}

Now we re-instate the cut-propagator on both sides of the equation by removing the delta function 
\begin{eqnarray}
\delta{(\ell^2-\mu^2)}\rightarrow\frac{i}{\ell^2-\mu^2},
\end{eqnarray}
to re-write the cut integral as a full Feynman integral.  To return from \eqref{eq:res4} to the Feynman integral representation 
we include a normalisation of $i$, and introduce $d(\ell)$ into the denominator, so that 
\begin{eqnarray}
A_4^{(1)}(1^+,2^+,3^+,4^+)&=&-i\frac{\spb1.2\spb3.4}{\spa1.2\spa3.4}\int \frac{d^{-2\epsilon}\mu}{(2\pi)^{-2\epsilon}}\int\frac{d^4\ell}{(2\pi)^4}\frac{i\mu^4}{d(\ell)d(\ell_1)d(\ell_2)d(\ell_3)}\nonumber\\
&=&\frac{i}{(4\pi)^{2-\epsilon}}\frac{\spb1.2\spb3.4}{\spa1.2\spa3.4}K_4.
\end{eqnarray}
We observe that we have correctly recovered the all-plus four-gluon amplitude 
given in \eqref{eq:allplus4} up to a factor of 2 which can naturally be attributed to the need to sum the two identical solutions associated with the two possible helicity assignments of the scalar particles.


\subsection{The all-plus five-gluon amplitude}

We now consider the slightly more complicated example of the all-plus five-glion one-loop amplitude. This was originally calculated to $\mc{O}(\epsilon^0)$ in \cite{Bern:1993mq} and to $\mc{O}(\epsilon)$ in \cite{Bern:1996ja}. Written in terms of master integrals it has the following form, 
\begin{eqnarray}
A_5^{(1)}(1^+,2^+,3^+,4^+,5^+)=\frac{i}{D_5}\frac{\epsilon(1-\epsilon)}{(4\pi)^{2-\epsilon}}\bigg[s_{23}s_{34} I_{4}^{(1),8-2\epsilon}+
s_{34}s_{45} I_{4}^{(2),8-2\epsilon}+s_{15}s_{45} I_{4}^{(3),8-2\epsilon}\nonumber\\+s_{15}s_{12} I_{4}^{(4),8-2\epsilon}+s_{12}s_{23} I_{4}^{(5),8-2\epsilon}+4i(4-2\epsilon)\epsilon(1234)I_5^{10-2\epsilon}\bigg],\nonumber\\
\end{eqnarray}
where $D_5=\spa1.2\spa2.3\spa3.4\spa4.5\spa5.1$ and $I_{4}^{(i),D}$ is a $D$-dimensional one-mass box, with the massive leg being formed from momenta $p_i$ and $p_{i-1}$. 
\begin{figure}
\begin{center}
\psfrag{o}{$1^+$}
\psfrag{m}{$5^+$}
\psfrag{A}{(a)}
\psfrag{B}{(b)}
\psfrag{C}{(c)}
\psfrag{D}{(d)}
\psfrag{E}{(e)}
\psfrag{F}{(f)}
\psfrag{G}{(g)}
\includegraphics[width=14cm]{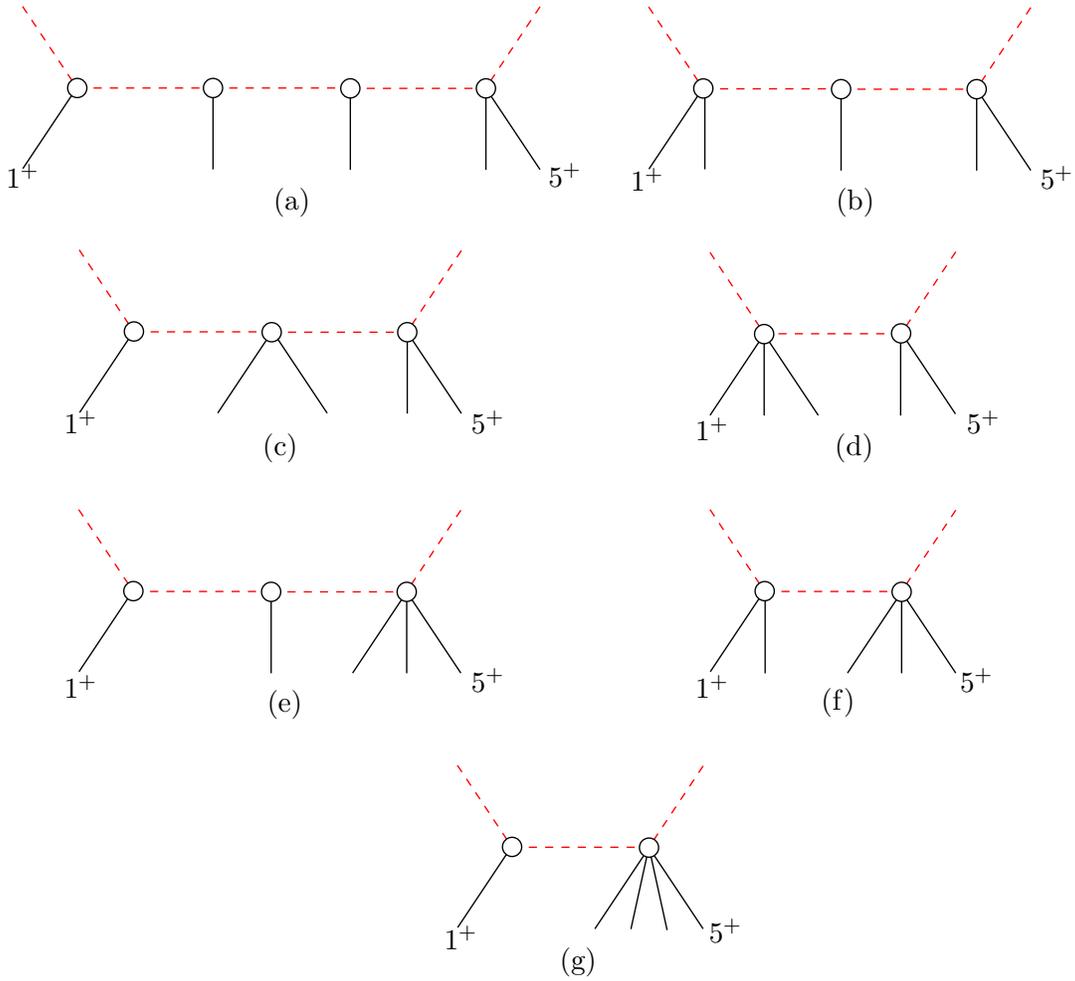}
\caption{
The MHV topologies associated with a $C_{1,5}$ cut of the all-plus five-gluon  amplitude. Here we have set $\n=p_5$ to eliminate any three-point vertex with containing
two scalars and  $p_5$. Other cuts are obtained from this one by cyclic relabeling of the external particles.}
\label{fig:5top}
\end{center}
\end{figure}

Since there are now five external particles, the one-loop basis functions 
include single-mass boxes and therefore we cannot calculate the full amplitude from one single cut. This is because the single mass box $I_{4}^{(i),D}$ cannot be detected in the $C_{i,i-1}$ cut. The possible topologies for the $C_{1,5}$ cut are shown in Fig.~\ref{fig:5top} where we have taken $\eta=p_5$ to eliminate any diagram with a three-point vertex containing $p_5$.  Each cut is separately gauge invariant, so that other cuts can be obtained by cyclic relabelling of the external particles.  

The spurious term structure of this amplitude is quite complicated, for a general  diagram we find three unique inverse factors of $\spab i.|\ell_{i-1}|.\eta$. The amplitude has a representation in terms of the one-loop scalar integrals, which are free of this sort of term. Therefore, we seek to remove these terms from our calculation. 

To begin this simplification, we first derive a relationship between the CSW vertex with a single leg off-shell and the three-point amplitude with all particles on-shell \cite{Badger:massrec}. 
We begin with the off-shell vertex of \cite{Boels:2007pj,Boels:2008ef}. 
\begin{figure}
\begin{center}
\psfrag{E}{$=$}
\psfrag{p}{$+$}
\psfrag{a}{$\alpha$}
\psfrag{b}{$\beta$}
\includegraphics[width=10cm]{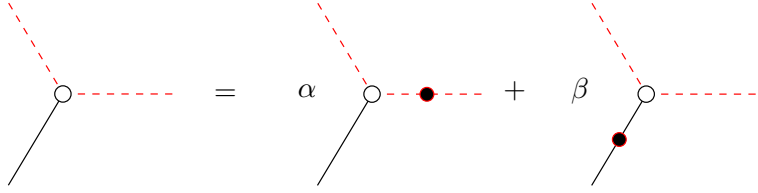}
\caption{The schematic structure of the three-point vertex with one massive scalar leg on-shell and one off-shell. A dotted leg represents the pinching of a propagator (scalar) or the absence of a spurious term associated with that particular gluon. When applied to a generic diagram, this identity leads to two contributions,  one term which pinches a propagator and a second term which has no spurious pole.}
\label{fig:onoff}
\end{center}
\end{figure}

\begin{figure}
\begin{center}
\psfrag{E}{$=$}
\psfrag{p}{$+$}
\psfrag{a}{$\alpha$}
\psfrag{b}{$\beta$}
\psfrag{c}{$\gamma$}
\includegraphics[width=14cm]{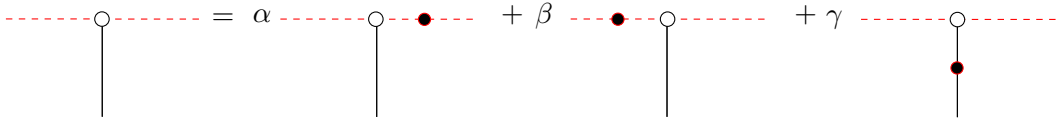}
\caption{The schematic structure of the three-point vertex with two off-shell scalars.  As in Fig.~5, dotted legs represents the pinching of a propagator (scalar) or the absence of a spurious term associated with that particular gluon. When applied to a generic diagram, this identity leads to three contributions,  two terms which pinch a propagator and a second term which has no spurious pole.}
\label{fig:onoffoff}
\end{center}
\end{figure}
\begin{eqnarray}
V_{CSW}(\ell,1^+,\ell_1)=-\frac{\mu^2\spbf \eta.|\ell.\ell_1|.\eta}{\spba \eta.|\ell|.1\spab 1.|\ell|.\eta}=-\frac{\mu^2\spb\eta.1}{\spab 1.|\ell|.\eta}
\end{eqnarray}
where $\ell$ and $p_1$ are on-shell and $\ell_1$ is off-shell. We can write the vertex in a more useful form, 
\begin{eqnarray}
\frac{\mu^2\spb \n.1}{\spone}=-\frac{d(\ell_1)(2\ell.\eta)}{\spab 1.|\ell|.\eta\spa1.\eta}-\frac{\spba 1.|\ell|.\eta}{\spa 1.\eta}.
\label{eq:ons}
\end{eqnarray}
When $\ell_1$ goes on-shell $d(\ell_1)=0$, and only the second term survives. However, in the cases we will consider, $\ell_1$ is an internal propagator and 
is always off-shell. 
The above expression is very useful since it relates the spurious parts of diagrams containing a three-point vertex to a diagram containing a four-point vertex (since the first term removes the propagator $d(\ell_1)$). The identity
\eqref{eq:ons} is shown diagramatically in Fig.~\ref{fig:onoff}. 
We also find the following more general expression for a three-point vertex with both scalar legs off-shell. 
\begin{eqnarray}
\frac{\mu^2\spb \n.i}{\spab i.|\ell_{i-1}|.\n}=-\frac{d(\ell_{i-1})\spab j.|\ell_{i}|.\n}{\spab i.|\ell_{i-1}|.\n\spa j.i}+\frac{d(\ell_i)\spab j.|\ell_{i-1}|.\n}{\spab i.|\ell_{i-1}|.\n\spa j.i}+\frac{\spab j.|\ell_{i-1}|.i}{\spa j.i} 
\label{eq:off}
\end{eqnarray}
which is illustrated in Fig.~\ref{fig:onoffoff}.

The notation we use in this calculation is as follows, we classify a generic contribution shown in Fig.~\ref{fig:5top} by its physical propagators $d(\ell_i)$ and its spurious poles $\spab i.|\ell_{i-1}|.\n$. For example the term ${a}_{123}$ has the denominators associated with Fig.~\ref{fig:5top}$(a)$ and spurious poles $\spone$, $\sptwo$ and $\spthree$.  Applying the CSW prescription to all the contributions in Fig.~\ref{fig:5top}, we find that
\begin{eqnarray}
a_{123}&=&-i\intOm[\mu^8].{d(\ell_1)d(\ell_2)d(\ell_3)d(\ell_4)}\frac{\spb\n.1\spb \n.2\spb \n.3\spb \n.4}{\spa 4.5\spone\spab 2.|\ell_1|.\eta\spab3.|\ell_2|.\n}\\
b_{123}&=& i\intOm[\mu^6].{d(\ell_2)d(\ell_3)d(\ell_4)}\frac{\spb \n.3\spb \n.4\spbf \n.|\ell.P_{1,2}|.\n}{\spa 1.2\spa 4.5\spab 1.|\ell|.\eta\spab 2.|\ell_1|.\eta\spab3.|\ell_2|.\n}\\
c_{123}&=&i\intOm[\mu^6].{d(\ell_1)d(\ell_2)d(\ell_4)}\frac{\spb\n.1\spb \n.2\spbf \n|.P_{1,2}.\ell|.\n}{\spa3.4\spa 4.5\spone\spab 2.|\ell_1|.\eta\spab3.|\ell_2|.\n}\\
d_{123}&=&-i\intOm[\mu^4].{d(\ell_2)d(\ell_4)}\frac{\spbf \n.|\ell.P_{1,2}|.\n\spbf \n|.P_{1,2}.\ell|.\n}{\spa1.2\spa3.4\spa 4.5\spab 1.|\ell|.\eta\spab 2.|\ell_1|.\eta\spab3.|\ell_2|.\n}\\
e_{123}&=&i\intOm[\mu^6].{d(\ell_1)d(\ell_3)d(\ell_4)}\frac{\spb\n.1\spbf \n.|\ell_{1}.P_{2,3}|.\n\spb \n.4}{\spa2.3\spa 4.5\spone\spab 2.|\ell_1|.\eta\spab3.|\ell_2|.\n}\\
f_{13}&=&i\intOm[\mu^4].{d(\ell_3)d(\ell_4)}\frac{\spbf \n.|\ell.P_{1,3}|.\n\spb \n.4}{\spa1.2\spa2.3\spa 4.5\spab 1.|\ell|.\eta\spab3.|\ell_2|.\n}\\
g_{2}&=&i\intOm[\mu^4].{d(\ell_1)d(\ell_4)}\frac{\spb \n.1^2}{\spa 2.3\spa3.4\spa4.5\spab 2.|\ell_1|.\eta}.
\end{eqnarray}
Applying eq.~\eqref{eq:ons} to these terms, we find that the spurious pieces containing $\spone$ generated from $a_{123}$, $c_{123}$ and $e_{123}$ cancel pairwise with the analogous contributions from $b_{123}$, $d_{123}$ and $f_{13}$. The remaining terms are free of $\spone$, i.e.
\begin{eqnarray}
a_{123}+b_{123} \rightarrow a_{23}+b_{23} \qquad c_{123}+d_{123} \rightarrow c_{23}+d_{23} \qquad e_{123}+f_{13} \rightarrow e_{23}+f_{23}.
\end{eqnarray}
With the various contributions given by the following,
 \begin{eqnarray}
 a_{23}&=&i\intOm[\mu^6].{d(\ell_1)d(\ell_2)d(\ell_3)d(\ell_4)}\frac{\spb \n.2\spb \n.3\spb \n.4\spba 1.|\ell|.\n}{\spa 1.\n\spa 4.5\spab 2.|\ell_1|.\eta\spab3.|\ell_2|.\n}\\
b_{23}&=&-i\intOm[\mu^6].{d(\ell_2)d(\ell_3)d(\ell_4)}\frac{\spb \n.3\spb \n.4\spba \n.|P_{1,2}|.\n}{\spa 1.\n\spa 1.2\spa 4.5\spab 2.|\ell_1|.\eta\spab3.|\ell_2|.\n}\\
c_{23}&= &-i\intOm[\mu^4].{d(\ell_1)d(\ell_2)d(\ell_4)}\frac{\spb \n.2\spbf \n|.P_{1,2}.\ell|.\n\spba 1.|\ell|.\n}{\spa 1.\n\spa3.4\spa 4.5\spab 2.|\ell_1|.\eta\spab3.|\ell_2|.\n}\\
d_{23}&=&i\intOm[\mu^4].{d(\ell_2)d(\ell_4)}\frac{\spbf \n|.P_{1,2}.\ell|.\n\spba \n.|P_{1,2}|.\n}{\spa 1.\n\spa1.2\spa3.4\spa 4.5\spab 2.|\ell_1|.\eta\spab3.|\ell_2|.\n}\\
e_{23}&=&-i\intOm[\mu^4].{d(\ell_1)d(\ell_3)d(\ell_4)}\frac{\spbf \n.|\ell_{1}.P_{2,3}|.\n\spb \n.4\spba 1.|\ell|.\n }{\spa1.\n\spa2.3\spa 4.5\spab 2.|\ell_1|.\eta\spab3.|\ell_2|.\n}\\
f_{23}&=&i\intOm[\mu^4].{d(\ell_3)d(\ell_4)}\frac{\spb \n.4(\spbf \n.|\ell.P_{1,3}|.\n\spa 2.\n-\spa1.2 \spb \n.1\spab \n.|\ell_3|.\n)}{\spa 1.\n\spa1.2\spa2.3\spa 4.5\spab 2.|\ell_1|.\eta\spab3.|\ell_2|.\n}.
\end{eqnarray}
Next we apply eq.~\eqref{eq:off} to $a_{23}$ with $i=3$ which generates three terms that combine with with $c_{23}$ and $e_{23}$ to remove $\spthree$.  Similarly, acting on $b_{23}$ with $j=2$ with eq.~\eqref{eq:off} produces terms that cancel the spurious $\spthree$ denominator in $d_{23}$ and $f_{23}$. So that
\begin{eqnarray}
a_{23}+c_{23}+e_{23} \rightarrow a_2+c_2+e_2  \qquad b_{23}+e_{23}+f_{23} \rightarrow b_2+e_2+f_2.
\end{eqnarray}
The remaining contributions are,
 \begin{eqnarray}
 a_{2}&=&i\intOm[\mu^4].{d(\ell_1)d(\ell_2)d(\ell_3)d(\ell_4)}\frac{\spb \n.2\spb \n.4\spba 1.|\ell|.\n\spba 3.|\ell_{1}|.2}{\spa 1.\n\spa2.3\spa 4.5\spab 2.|\ell_1|.\eta}\\
b_{2}&=&-i\intOm[\mu^4].{d(\ell_2)d(\ell_3)d(\ell_4)}\frac{\spb \n.4\spba \n.|P_{1,2}|.\n\spba 3.|\ell_{1}|.2}{\spa 1.\n\spa 1.2\spa2.3\spa 4.5\spab 2.|\ell_1|.\eta}\\
c_{2}&=&i\intOm[\mu^4].{d(\ell_1)d(\ell_2)d(\ell_4)}\frac{\spb \n.2\spa1.2\spb\n.1\spba 1.|\ell|.\n}{\spa 1.\n\spa2.3\spa3.4\spa 4.5\spab 2.|\ell_1|.\eta}\\
d_{2}&=&-i\intOm[\mu^4].{d(\ell_2)d(\ell_4)}\frac{\spba \n.|P_{1,2}|.\n\spb\n.1}{\spa 1.\n\spa2.3\spa3.4\spa 4.5\spab 2.|\ell_1|.\eta}\\
e_{2}&=&-i\intOm[\mu^4].{d(\ell_1)d(\ell_3)d(\ell_4)}\frac{\spb\n.3\spb \n.4\spba 1.|\ell|.\n}{\spa 1.\n\spa2.3\spa 4.5\spab 2.|\ell_1|.\eta}\\
f_{2}&=&i\intOm[\mu^4].{d(\ell_3)d(\ell_4)}\frac{\spb \n.4\spa 2.\n \spb 3.\n}{\spa 1.\n\spa 1.2\spa2.3\spa 4.5\spab 2.|\ell_1|.\eta}\\
g_{2}&=&i\intOm[\mu^4].{d(\ell_1)d(\ell_4)}\frac{\spb \n.1^2}{\spa 2.3\spa3.4\spa4.5\spab 2.|\ell_1|.\eta}.
\end{eqnarray}
Finally,  putting $a_2$, $b_2$, $e_2$ and $f_2$ over a common denominator we find, 
\begin{eqnarray}
a_2+b_2+e_2+f_2=a+b.
\end{eqnarray}
Similarly, when $c_2$, $d_2$ and $g_2$ are combined we find that
\begin{eqnarray}
c_2+d_2+g_2=c,
\end{eqnarray}
where the three remaining terms are free of spurious singularities, 
\begin{eqnarray}
a&=&-i\intOm[\mu^4].{d(\ell_1)d(\ell_2)d(\ell_3)d(\ell_4)}\frac{\spbsev \n.4.3.2.1.\ell.\n}{D_5}\\
b&=&i\intOm[\mu^4].{d(\ell_2)d(\ell_3)d(\ell_4)} \frac{s_{34}s_{4\n}}{D_5}\\
c&=& i\intOm[\mu^4].{d(\ell_1)d(\ell_2)d(\ell_4)}\frac{s_{12}s_{1\n}}{D_5}
\end{eqnarray}
We note that $b$ and $c$ are scalar integrals and require no further
simplification. However, $a$ has a linear dependance on $\ell$ and requires some
further manipulation.  We use momentum conservation to re-write the numerator as
a sum of two traces 
\begin{eqnarray}
a=i\intOm[\mu^4].{d(\ell_1)d(\ell_2)d(\ell_3)d(\ell_4)}\frac{s_{4\n}\trp (\n21\ell)+s_{12}\trp(\n41\ell)}{D_5}
\end{eqnarray}
where $\trp(a,b,c,d)= \spb a.b\spa b.c\spb c.d\spa d.a$.  
Explicit evaluation of these traces leads to a sum of scalar boxes and a linear pentagon,
\begin{eqnarray}
a=i\intOm[\mu^4].{d(\ell_1)d(\ell_2)d(\ell_3)d(\ell_4)}\frac{1}{2D_5}\bigg(-d(\ell_1)s_{34}s_{4\n}+d(\ell_2)s_{1\n}s_{4\n}-d(\ell_3)s_{12}s_{1\n}\nonumber\\
+d(\ell_4)s_{12}s_{23}-2i\bigg(s_{4\n}\epsilon(12\n\ell)+s_{12}\epsilon(14\n\ell)\bigg)\bigg).
\end{eqnarray}
It is possible to simplify the linear pentagon using the following identities 
\begin{eqnarray}
s_{4\n}\epsilon(12\n\ell)
&=&(d(\ell_3)-d(\ell_4))\epsilon(12\n\ell)-\mu^2\epsilon(214\n)+d(\ell_1)\epsilon(24\ell\n)\nonumber\\&&+(\spbsev 2.\ell.1.4.\ell.\n.2-\spasev 2.\ell.1.4.\ell.\n.2),\\
s_{12}\epsilon(14\n\ell)&=&(d(\ell_2)-d(\ell_1))\epsilon(14\n\ell)+\mu^2\epsilon(14\n\ell)+d(\ell_4)\epsilon(14\n\ell)\nonumber\\
&&-\spbsev 1.4.\ell.\n.2.\ell.1+\spasev 1.4.\ell.\n.2.\ell.1,
\end{eqnarray}
so that,
\begin{eqnarray}
s_{12}\epsilon(14\n\ell)+s_{4\n}\epsilon(12\n\ell)=-d(\ell_1)\epsilon(34\n\ell)+d(\ell_2)\epsilon(14\n\ell)+d(\ell_3)\epsilon(12\n\ell)\nonumber\\-d(\ell_4)\epsilon(123\ell)+2\mu^2\epsilon(14\n2).
\end{eqnarray}
Here every $\epsilon$ tensor which multiplies a denominator is associated with a linear box. However, the remaining three momenta in the tensor are precisely the three massless momenta of each box, which are the basis vectors for the Passarino-Veltman (PV) expansion. Therefore, each time the $\epsilon$ tensor is contracted with a basis vector from the PV expansion, there is a repeated momenta in the $\epsilon$ tensor and the term vanishes. We are thus left with only the last term, which is associated with a scalar pentagon.

The total contribution to the $C_{1,5}$ single-cut is thus given by $a+b+c$ 
which is the following sum of scalar boxes and a scalar pentagon
\begin{eqnarray}
A^{(1)}_5(1^+,2^+,3^+,4^+,5^+)_{C_{1,5}~cut}&=&i\intOm[\mu^4].{d(\ell_1)d(\ell_2)d(\ell_3)d(\ell_4)}\frac{1}{2D_5}\bigg(d(\ell_1)s_{34}s_{4\n}+d(\ell_2)s_{1\n}s_{4\n}\nonumber\\
&&+d(\ell_3)s_{12}s_{1\n}+d(\ell_4)s_{12}s_{23}-4i\mu^2\epsilon(1234)\bigg).
\end{eqnarray} 
If we re-instate the cut-propagator then we find that,
\begin{eqnarray}
A^{(1)}_5(1^+,2^+,3^+,4^+,5^+)&=&-\frac{i}{2(4\pi)^{2-\epsilon}D_5}\bigg(s_{34}s_{45}I_4^{(2)4-2\epsilon}[\mu^4]+s_{51}s_{45}I_4^{(3)4-2\epsilon}[\mu^4]\nonumber\\&&+s_{12}s_{51}I_4^{(4)4-2\epsilon}[\mu^4]+s_{12}s_{23}I_4^{(5)4-2\epsilon}[\mu^4]+4i\epsilon(1234)I_5^{4-2\epsilon}[\mu^6]\bigg).\nonumber\\
\end{eqnarray}
Up to the usual factor of two, we observe that the above equation correctly reproduces the known result, apart from the term proportional to $I_4^{(5)4-2\epsilon}$. Of course, this is completely expected since the missing box has momenta $p_1$ and $p_5$ emitted from a single vertex, which cannot be found from a single cut in the $C_{1,5}$ channel we have calculated. 
Therefore to obtain the missing term we could take a cut in a different channel. Alternatively, we can use the high degree of symmetry of the amplitude under a cyclic permutation of gluons to generate the correct coefficient of this term: $s_{23}s_{34}$.


\section{Amplitudes containing gluons of negative helicity}

In the following section, we turn our attention to the calculation of four-point amplitudes involving mixtures of negative and positive helicity gluons. The first helicity configuration we consider is the mostly-plus amplitude with a single negative helicity gluon. In supersymmetric theories this amplitude vanishes, so the rational contribution in QCD is obtained from computing the  scalar loop $\mc{N}=0$ contribution in $D$ dimensions. As a further example, we also  calculate the scalar loop contributions to the four-gluon MHV amplitude $A^{(1),\mc{N}=0}(1^-,2^-,3^+,4^+)$. Finally, and as an example of using the single-cut in four dimensions,  we calculate the amplitude $n$-gluon MHV amplitude for the specific helicity configuration where the two negative helicity gluons are adjacent in $\mc{N}=4$ SYM.

\subsection{The mostly-plus four-gluon amplitude} 

The mostly-plus four-gluon amplitude has been calculated to $\mc{O}(\epsilon)$ in \cite{Bern:1995db}, and has the following form, 
\begin{eqnarray}
\label{eq:resmppp}
A^{(1)}_4(1^-,2^+,3^+,4^+)=\frac{2i}{(4\pi)^{2-\epsilon}}\frac{\spb 2.4^2}{\spb1.2\spa2.3\spa3.4\spb4.1}\frac{st}{u}\bigg(K_4+\frac{st}{2u}J_4\nonumber\\+\frac{t(u-s)}{su}J_3(s)+\frac{s(u-t)}{tu}J_3(t)-\frac{t-u}{s^2}J_2(s)-\frac{s-u}{t^2}J_2(t)\bigg).
\end{eqnarray}
Here we have introduced the notation used in~\cite{Brandhuber:2005jw}
\begin{eqnarray}
J_n(s)=I_n[\mu^2](s)=(-\epsilon)I_n^{6-2\epsilon}(s)
\end{eqnarray}
for the massless bubble and one-mass triangle integrals.

\begin{figure} 
\begin{center}
\psfrag{o}{$1^-$}
\psfrag{n}{$4^+$}
\includegraphics[width=8cm]{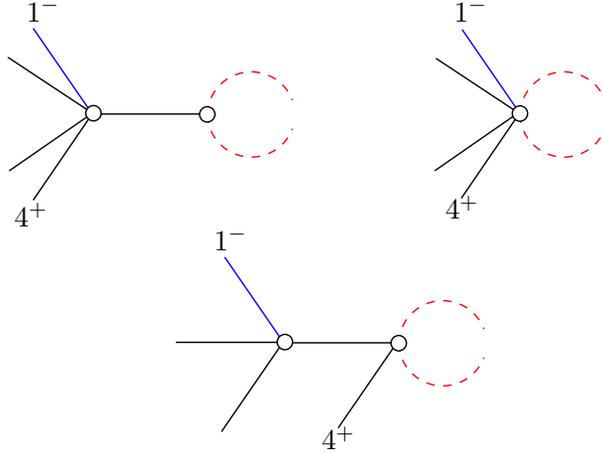}
\caption{Topologies which contain the two cut scalar particles at the same vertex and are free of spurious terms. Each is proportional to either a massless bubble or a tadpole diagram and as a result do not contribute to the amplitude.}
\label{fig:41b}
\end{center}
\end{figure}

\begin{figure}
\begin{center}
\psfrag{o}{$1^-$}
\psfrag{n}{$4^+$}
\psfrag{A}{(a)}
\psfrag{B}{(b)}
\psfrag{C}{(c)}
\psfrag{D}{(d)}
\psfrag{E}{(e)}
\includegraphics[width=11cm,height=11cm]{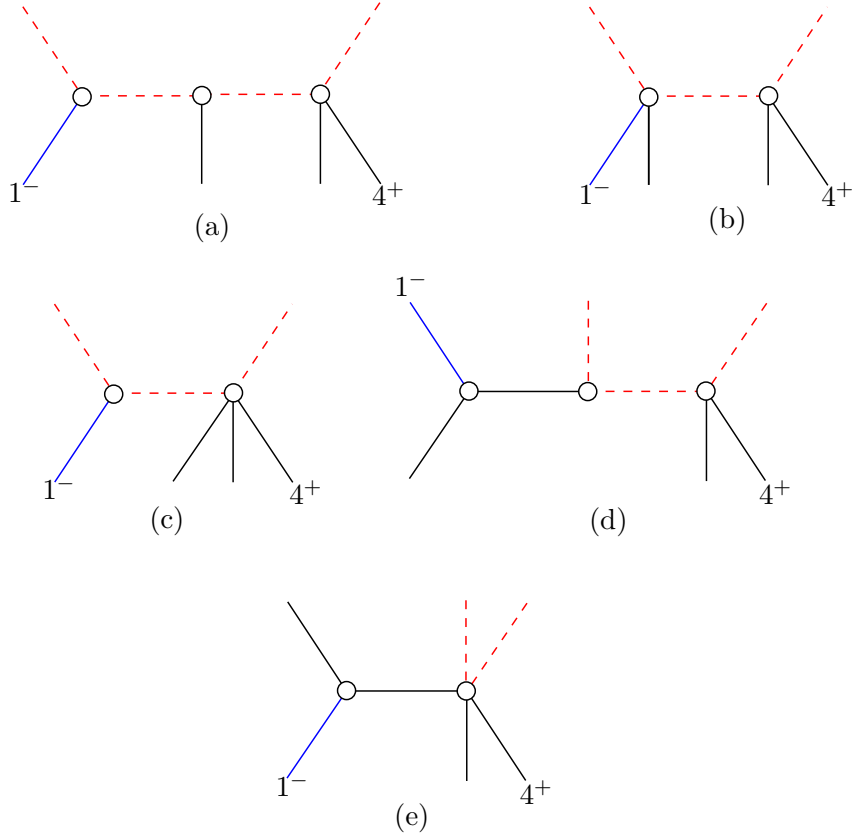}
\caption{Topologies which contribute to $A_4^{(1)}(-,+,+,+)$ in the 
$C_{1,4}$ cut. Here we have set $\n=4$. Gluons of negative helicity are shown in blue The diagrams shown above contain spurious terms which must be removed.}
\label{fig:41a}
\end{center}
\end{figure}

As in Sec.~\ref{sec:allplus4} we choose to make the $C_{1,4}$ cut and set 
$\n=4$.   The contributing diagrams are shown in Fig.~\ref{fig:41b} and Fig.~\ref{fig:41a}. The diagrams of Fig.~\ref{fig:41a}
are very similar to those previously encountered, however, Fig.~\ref{fig:41b} shows a new class of diagrams associated with vertices containing both of the massive scalars. Although we are free to choose $\n$ to be any massless vector the choice $\n=p_4$ is sensible since it automatically removes any spurious singularities from these graphs.
With this choice of $\n$, the diagrams in Fig.~\ref{fig:41b} are simply proportional to massless no-scale bubble and tadpole integrals, which in dimensional regularisation vanish.   

The five remaining diagrams shown in Fig.~\ref{fig:41a} are given by,
\begin{eqnarray}
a_2&=&-i\intOm[\mu^4].{d(\ell_1)d(\ell_2)d(\ell_3)}\frac{\spab 1.|\ell|.\n\spb\n.3\spb\n.2}{\spb1.\n\spa3.4\sptwo}\\
b_{23}&=&-i\intOm[\mu^2].{d(\ell_2)d(\ell_3)}\frac{\spab 1.|\ell|.\n\spab 1.|\ell_2|.\n^2}{\spa1.2\spa3.4\sptwo\spthree}\\
c_{2}&=&-i\intOm[\mu^2].{d(\ell_1)d(\ell_3)}\frac{\spab 1.|\ell|.\n^2}{\spa2.3\spa3.4\sptwo}\\
d_3&=&i\intOm[\mu^4].{d(\ell_2)d(\ell_3)}\frac{\spb \n.2^3}{\spb1.2\spa3.4\spb1.\n\spthree}\\
e_3&=&i\intOm[\mu^2].{d(\ell_3)}\frac{\spb \n.2^3\spbf\n.|\ell.\ell|.\n}{\spb1.2\spa3.4\spbf\n.|P_{12}.3|.\n\spb1.\n\spthree}.
\end{eqnarray}

Although Fig.~\ref{fig:41a}$(e)$ contains both cut scalars at the same vertex, it cannot be neglected in the same manner as the diagrams shown in Fig.~\ref{fig:41b}. This is because of the presence of the spurious term $\spthree$, hence the integral is not identifiable as a loop integral
in the physical basis. The inclusion of this diagram is essential in the removal of the spurious terms associated with Fig.~\ref{fig:41a}$(d)$.

We use eq.~\eqref{eq:off} with $i=2$ and $j=\n$ to combine $a_2$ with $b_{23}$ and $c_2$, 
\begin{equation}
a_2+b_{23} + c_2 \to a + b + b_3 + c,
\end{equation}
where the remaining terms are free of the spurious pole $\sptwo$,
\begin{eqnarray}
a&=&-i\intOm[\mu^2].{d(\ell_1)d(\ell_2)d(\ell_3)}\frac{\spab 1.|\ell|.\n\spab\n.|\lo|.2\spb\n.3}{\spb1.\n\spa3.4\spa \n.2}\\
b&=&-i\intOm[\mu^2].{d(\ell_2)d(\ell_3)}\frac{\spab 1.|\ell|.\n\spa\n.1\spb \n.3}{\spa1.2\spa3.4\spb1.\n\spa\n.2}\\
b_3&=&-i\intOm[\mu^2].{d(\ell_2)d(\ell_3)}\frac{\spab 1.|\ell|.\n\spab 1.|\ell_2|.\n\spb\n.2}{\spa1.2\spa3.4\spb1.\n\spthree}\\
c&=&i\intOm[\mu^2].{d(\ell_1)d(\ell_3)}\frac{s_{1\n}\spab 1.|\ell|.\n}{\spb1.\n\spa\n.2\spa2.3\spa3.4}.
\end{eqnarray}
Next we remove the spurious pole $\spthree$ by using eq.~\eqref{eq:off} with $i=3$ and $j=\n$ in $d_3$ and combining the resulting terms with $b_3$ and $e_3$, so that
\begin{equation}
d_3 + b_3 +e_3 \to d + e + b^{(1)} + b^{(2)},
\end{equation}
where 
\begin{eqnarray}
d&=&i\intOm[\mu^2].{d(\ell_2)d(\ell_3)}\frac{\spb \n.2^3\spab 3.|\ell|.\n}{\spb1.2\spa3.4\spb\n.3\spa\n.3\spb1.\n}\\
e&=&0 \\
b^{(1)}&=&i\intOm[\mu^2].{d(\ell_2)d(\ell_3)}\frac{\spb\n.2\spab1.|\ell|.\n\spa\n.1}{\spa1.2\spa3.4\spb1.\n\spa\n.3}\\
b^{(2)}&=&i\intOm[\mu^2].{d(\ell_2)d(\ell_3)}\frac{\spb\n.2^2\spab\n.|\lt|.\n\spb2.3}{\spb1.2\spa3.4\spb1.\n\spa\n.3\spb\n.3}.
\end{eqnarray}

All remaining contributions are now free of spurious poles and we can straightforwardly perform tensor reduction. We note that $d$ and $c$ vanish upon tensor reduction, leaving $a$, $b$ and $b^{(1),(2)}$.  The most complicated term is $a$ and this reduces to,
\begin{eqnarray}
a=\frac{i}{(4\pi)^{2-\epsilon}}\frac{\spb2.4^2}{\spb1.2\spa2.3\spa3.4\spb4.1}\frac{st}{u}\bigg(K_4+\frac{st}{2u}J_4+\frac{t}{2u}J_3(t)-\bigg(1+\frac{t}{u}\bigg)J_3(s)\bigg),
\end{eqnarray}
where we have reinstated the cut propagator to write this contribution in terms of the basis integrals.  
We note that with the $C_{1,4}$ cut, we expect to reconstruct $J_3(s_{23})$ triangle integrals (i.e. one mass integrals with its two on-shell legs given by 
$p_1$ and $p_4$) but not $J_3(s_{14})$ integrals where the off-shell leg is $(p_1+p_4)$. Of course, momentum conservation ensures that $J_3(s_{23})= J_3(s_{14})$ so we only partially reconstruct the coefficient of $J_3(t)$.
We therefore drop all of the $J_3(t)$ terms we find in this cut. 
The remaining contributions are,
\begin{eqnarray}
b+b^{(1)}&=&-\frac{i}{(4\pi)^{2-\epsilon}}\frac{\spb2.4^2}{\spb1.2\spa2.3\spa3.4\spb4.1}\frac{t^2}{su}J_2(s)\nonumber\\
b^{(2)}&=&-\frac{i}{(4\pi)^{2-\epsilon}}\frac{\spb2.4^2}{\spb1.2\spa2.3\spa3.4\spb4.1}\bigg(tJ_3(s)-\frac{t}{s}J_2(s)\bigg).
\end{eqnarray}
When we sum the various contributions, we recover all of the terms in the amplitude which depend on $s$, 
\begin{eqnarray}
A^{(1)}(1^-,2^+,3^+,4^+)_{41~ cut}=\frac{i}{(4\pi)^{2-\epsilon}}\frac{\spb2.4^2}{\spb1.2\spa2.3\spa3.4\spb4.1}\frac{st}{u}\bigg(K_4+\frac{st}{2u}J_4\nonumber\\
+\frac{t(s-u)}{su}J_3(s)+\frac{u-t}{s^2}J_2(s)\bigg)
 \end{eqnarray}
To reconstruct the remaining coefficients one could perform either a $C_{2,1}$ cut or use the symmetry of the amplitude under a $2\leftrightarrow 4 \equiv t\leftrightarrow s$ exchange,
\begin{eqnarray}
A^{(1)}(1^-,2^+,3^+,4^+)_{12~ cut}=\frac{i}{(4\pi)^{2-\epsilon}}\frac{\spb2.4^2}{\spb1.4\spa4.3\spa3.2\spb2.1}\frac{st}{u}\bigg(K_4+\frac{st}{2u}J_4\nonumber\\
+\frac{s(t-u)}{st}J_3(t)+\frac{u-s}{t^2}J_2(t)\bigg).
 \end{eqnarray}
Combining the distinct terms from the $C_{1,4}$ and $C_{2,1}$ cuts, and adding a factor two for the two possible helicity assignments of the scalar particles,
we recover the known result of eq.~\eqref{eq:resmppp}.

\subsection{The scalar-loop contribution to a four-gluon MHV amplitude}

\begin{figure}
\begin{center}
\psfrag{n}{$3^+$}
\psfrag{o}{$4^+$}
\psfrag{A}{(a)}
\psfrag{B}{(b)}
\psfrag{C}{(c)}
\includegraphics[width=14cm]{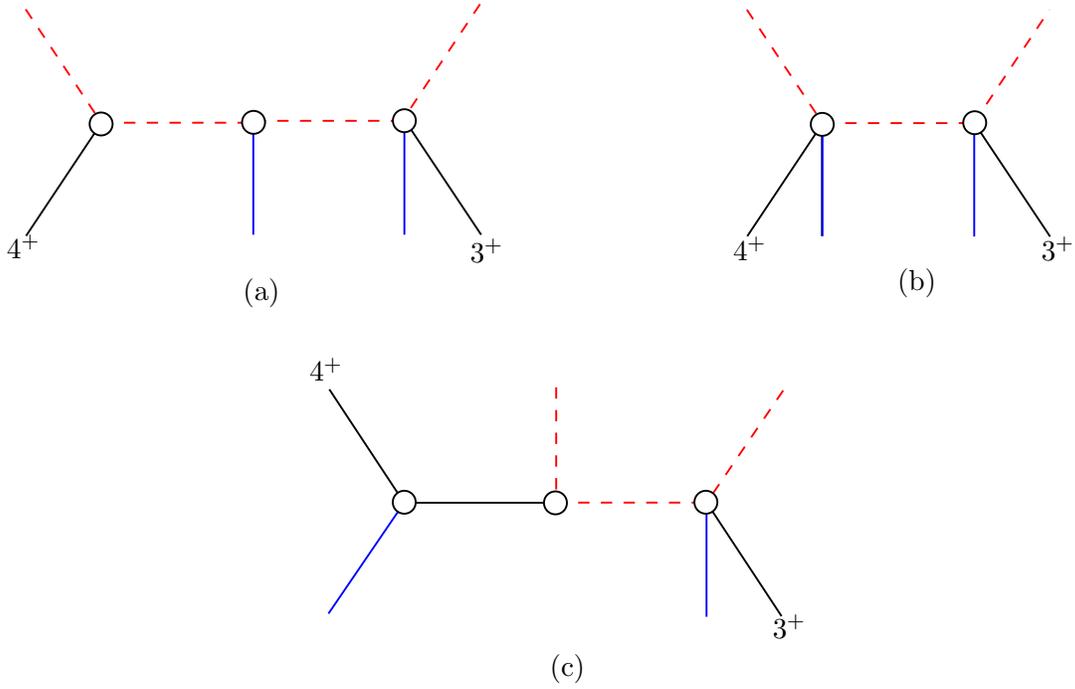}
\caption{Topologies which contribute to the scalar loop amplitude $A_n^{(1,\mc{N}=1)}(-,-,+,+)$ in the $C_{4,3}$ cut. Here we have set $\n=3$. The two external gluons with negative helicity are shown in blue. }
\label{fig:MHV}
\end{center}
\end{figure}

As a final application of the single cut in $D$ dimensions,  we calculate the scalar-loop contribution to the four-gluon MHV amplitude with adjacent negative helicity gluons. This amplitude corresponds to the result in $\mc{N}=0$ supersymmetric QCD and 
has been given to $\mc{O}(\epsilon)$ in \cite{Bern:1995db}
\begin{eqnarray}
\label{eq:4ptMHV}
A_4^{(1),\mc{N}=0}(1^-,2^-,3^+,4^+)&=&2\frac{A^{(0)}_4}{(4\pi)^{2-\epsilon}}\bigg(-\frac{t}{s}K_4+\frac{1}{s}J_2(t)+\frac{1}{t}I_2^{D=6-2\epsilon}(t)\bigg),
\end{eqnarray}
where 
\begin{eqnarray}
A^{(0)}_4=i\frac{\spa1.2^3}{\spa2.3\spa3.4\spa4.1}.
\end{eqnarray}

The amplitude can be fully determined from the $C_{4,3}$ cut.  With the choice $\n=p_3$ there are three contributing topologies which are shown in Fig.~\ref{fig:MHV}.  The only spurious terms are of the form $\spfour$ such that,
\begin{eqnarray}
a_4&=&i\intOm[\mu^2].{d(\ell_4)d(\ell_{41})d(\ell_{42})}\frac{\spab 1.|\ell_4|.\n\spab 2.|\ell|.\n\spab2.|\ell_{41}|.\n\spb 4.\n}{\spa 2.3\spb1.\n\spb2.\n\spab4.|\ell|.\n}\nonumber\\
b_4&=&i\intOm[1].{d(\ell_{41})d(\ell_{42})}\frac{\spab 1.|\ell|.\n^2\spab 1.|\ell_4|.\n\spab2.|\ell_{41}|.\n}{\spa1.4\spa2.3\spb2.\n^2\spab4.|\ell|.\n}\nonumber\\
c&=&-i\intOm[\mu^2].{d(\ell_{41})d(\ell_{42})}\frac{\spb4.\n^3\spba\n.|\ell|.2}{\spa2.3\spb4.1\spb2.\n^2\spb\n.1},
\end{eqnarray}
where we have introduced the generalised notation
\begin{equation}
\ell_{ij}=\ell-\sum_{k=i}^{j}p_k.
\end{equation}
Applying eq.~\eqref{eq:ons}
to $a_4$ and combining with $b_4$, 
the dependence on $\spab 4.|\ell|.\n$ is eliminated, so that
\begin{equation}
a_4+b_4 \to a+b,
\end{equation} 
with
\begin{eqnarray}
a&=&i\intOm[1].{d(\ell_4)d(\ell_{41})d(\ell_{42})}\frac{\spab 1.|\ell_4|.\n\spab 2.|\ell|.\n\spab2.|\ell_{41}|.\n\spab\n.|\ell|.4}{\spa4.\n\spa 2.3\spb1.\n\spb2.\n}\\
b&=&i\intOm[1].{d(\ell_{41})d(\ell_{42})}\frac{\spa1.\n\spab1.|\ell|.\n\spab 1.|\ell_4|.\n\spab2.|\ell_{41}|.\n}{\spa1.4\spa4.\n\spa2.3\spb2.\n^2}\nonumber\\&&\qquad \qquad+
i\intOm[1].{d(\ell_{41})}\frac{\spab 1.|\ell_4|.\n\spab2.|\ell_{41}|.\n\spb4.\n}{\spb1.\n\spa4.\n\spa2.3\spb2.\n^2}.
\end{eqnarray}
We observe that $b$ and $c$ vanish upon Passarino-Veltman reduction 
while $a$ becomes\footnote{We follow \cite{Brandhuber:2005jw} and drop the terms proportional to tadpoles and massless bubbles}, 
\begin{eqnarray}
a&=&-A^{(0)}_4\bigg(-\frac{t}{s}\intOm[\mu^4].{d(\ell_4)d(\ell_{41})d(\ell_{42})}+\intOm[1].{d(\ell_{41})}\bigg(\frac{\mu^2}{s}-\frac{2\mu^2}{3t}+\frac{1}{6}\bigg)\bigg).
 \end{eqnarray}
Reinstating the cut-propagator (and the normalisation $i$) we find that the contribution to the scalar-loop four-gluon amplitude from the $C_{4,3}$ cut is given by,
\begin{eqnarray}
A^{(1)\mc{N}=0}(1^-,2^-,3^+,4^+)_{43-cut}=-\frac{A^{(0)}_4}{(4\pi)^{2-\epsilon}}\bigg(\frac{t}{s}K_4-\bigg(\frac{1}{s}-\frac{2}{3t}\bigg)J_2(t)-\frac{1}{6}I_2(t)\bigg)
\end{eqnarray}
This can be recast in terms of the $D=6-2\epsilon$ bubble integral using,
\begin{eqnarray}
I_2(t)=\frac{4}{t}J_2(t)+\frac{6}{t}I_2^{D=6-2\epsilon}(t)
\end{eqnarray}
to recover eq.~\eqref{eq:4ptMHV} up to the usual factor of 2 which we again associate with the two degenerate ways of assigning the helicity to the scalar.

\subsection{An $n$-gluon one-loop MHV amplitude in $\mc{N}=4$ SYM}

To show the versatility of the single cut method, we use it to calculate the $n$-gluon MHV amplitudes in $\mc{N}=4$ SYM with adjacent negative helicity gluons. Since these amplitudes 
are completely cut-constructible, we use four-dimensional massless CSW rules to construct the tree-level inputs. The analytic result is well known~\cite{BDDK:uni1} and is independent of the position of the two negative helicity gluons. It has the following form, 
\begin{equation}
\label{eq:neq4known}
A^{\mc{N}=4~MHV}_{n}=c_{\Gamma}A^{(0)}_n V^g_n,
\end{equation}
where 
\begin{equation}
c_{\Gamma}=\frac{1}{(4\pi)^{2-\epsilon}}\frac{\Gamma(1+\epsilon)\Gamma^2(1-\epsilon)}{\Gamma(1-2\epsilon)}
\end{equation}
and $V_n^g$ is a sum of only two-mass easy and one-mass box functions, $\Ftme$ and $\Fom$, respectively, 
\begin{eqnarray}
V_n^g=\sum_{i=1}^n\Fom(s_{i,i+2};s_{i,i+1},s_{i+1,i+2})
+\frac{1}{2}\sum_{i=1}^{n}\sum_{j=i+3}^{n+i-3}
\Ftme(s_{i,j},s_{i+1,j-1};s_{i+1,j},s_{i,j-1}).\nonumber\\
\end{eqnarray}
Here $\Ftme$ and $\Fom$ are related to the master scalar box integrals via,
\begin{eqnarray}
I_{4}^{1m}(s_{i,i+2};s_{i,i+1},s_{i+1,i+2})
&=&c_{\Gamma}\frac{-2\Fom(s_{i,i+2};s_{i,i+1},s_{i+1,i+2})}{s_{i,i+1}s_{i+1,i+2}}\nonumber\\
I_{4}^{2me}(s_{i,j},s_{i+1,j-1};s_{i+1,j},s_{i,j-1})&=&c_{\Gamma}\frac{-2\Ftme(s_{i,j},s_{i+1,j-1};s_{i+1,j},s_{i,j-1})}{s_{i+1,j}s_{i,j-1}-s_{i,j}s_{i+1,j+1}}.
\end{eqnarray}

\begin{figure}
\begin{center}
\psfrag{n}{$(i-1)^+$}
\psfrag{o}{$i^+$}
\psfrag{j}{$j$}
\psfrag{A}{(a)}
\psfrag{B}{(b)}
\psfrag{C}{(c)}
\includegraphics[height=8cm]{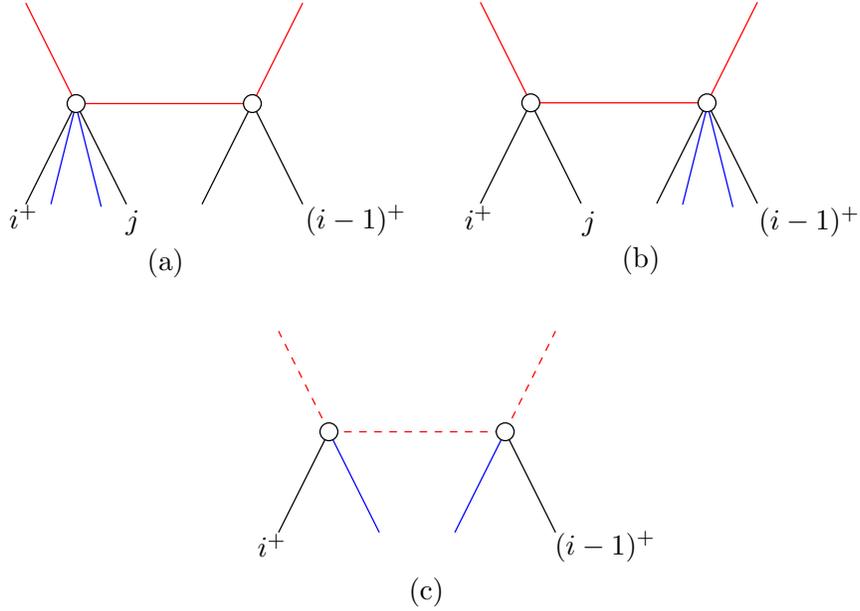}
\caption{Topologies which contribute to the 
$\mc{N}=4$ amplitude $A_n^{(1,\mc{N}=4)}(-,-,+,\ldots,+)$ in the $C_{i(i-1)}$ cut. The two external gluons with negative helicity are shown in blue. Particles which depend on the loop momenta are red, dashed lines indicate that the species is not fixed, solid lines represent gluons. }
\label{fig:N4}
\end{center}
\end{figure}

We consider the $C_{i(i-1)}$ cut where $i$ and $(i-1)$ are positive helicity gluons.
For this cut there are three allowed classes of diagram shown in Fig.~\ref{fig:N4}. If we choose $\n=p_{i-1}$, then the allowed range of $j$  for fixed $i$ is, 
\begin{eqnarray}
2\le j \le i-3 && {\rm for~Fig.}~\ref{fig:N4}(a), \\
i\le j \le n && {\rm for~Fig.}~\ref{fig:N4}(b).
\end{eqnarray}
In Fig.~\ref{fig:N4}$(a)$ and $(b)$, the species of the loop particle is 
fixed by angular momentum to be gluonic while for Fig.~\ref{fig:N4}$(c)$,  there is no such restriction and the entire multiplet is allowed to propagate in the loop. 
We first consider Fig.~\ref{fig:N4}$(a)$ which is given by,
\begin{eqnarray}
\mc{D}^{(a)}_{i,i-1}=\sum_{j=2}^{i-3}\mc{D}^{a,j}_{i,i-1},
\label{eq:sum}
\end{eqnarray}
with
\begin{eqnarray}
\mc{D}^{a,j}_{i,i-1}=-\int \frac{d^4\ell}{(2\pi)^4}\delta(\ell^2)\frac{A^{(0)}_n\spa \im.i\spa j.\jp\spab \ell.|\ell_{i,j}|.\n^2}{d(\ell_{i,j})\spa \ell.i\spab j.|\ell_{i,j}|.\n\spab \jp.|\ell_{i,j}|.\n\spa\im.\ell},
\end{eqnarray}
where $\ell_{i,j}=\ell-P_{i,j}$.  We use the Schouten identity to write
\begin{eqnarray} 
\mc{D}^{a,j}_{i,i-1}=-A^{(0)}_n\int \frac{d^4\ell}{(2\pi)^4}\frac{\spa \im.i}{\spa \ell.i\spa\im.\ell}\bigg(\mc{G}^i_j(P_{i,j})-\mc{G}^i_{\jp}(P_{i,j})\bigg)
\end{eqnarray}
with 
\begin{eqnarray}
\mc{G}^i_j(P_{a,b})=\frac{\spa j.\ell\spab\ell.|\ell_{a,b}|.\n}{d(\ell_{a,b})\spab j.|\ell_{a,b}|.\n}.
\end{eqnarray}
Inserting this form into eq.~\eqref{eq:sum}, we find 
\begin{eqnarray}
\mc{D}^{(a)}_{i,i-1}=-A^{(0)}_n\int \frac{d^4\ell}{(2\pi)^4}\delta(\ell^2)\frac{\spa \im.i}{\spa \ell.i\spa\im.\ell}\bigg(\sum_{j=2}^{i-3}\mc{G}^i_j(P_{i,j})
-\sum_{j=2}^{i-3}\mc{G}^i_{\jp}(P_{i,j})\bigg).
\end{eqnarray}
Making the shift $j \rightarrow j-1$ in the second sum, we find,
\begin{eqnarray}
\mc{D}^{(a)}_{i,i-1}&=&-A^{(0)}_n\int \frac{d^4\ell}{(2\pi)^4}\delta(\ell^2)\frac{\spa \im.i}{\spa \ell.i\spa\im.\ell}\bigg(\sum_{j=3}^{i-3}\bigg(\mc{G}^i_j(P_{i,j})-\mc{G}^i_{j}(P_{i,j-1})\bigg)\nonumber\\&&+\mc{G}^i_2(P_{i,2})-\mc{G}^i_{i-2}(P_{i,i-3})\bigg).
\end{eqnarray}
We will now show that $\mc{G}^i_j(P_{i,j})-\mc{G}^i_{j}(P_{i,j-1})$ and $\mc{G}^i_{i-2}(P_{i,i-3})$ are free of spurious singularities. The remaining term $\mc{G}^i_2(P_{i,2})$ still contains a spurious term, which we will eventually cancel against contributions from Fig.~\ref{fig:N4}$(c)$. We begin with $\mc{G}^i_{i-2}(P_{i,i-3})$ which has the form, 
\begin{eqnarray}
\mc{G}^i_{i-2}(P_{i,i-3})=\frac{\spa {(i-2)}.\ell\spab\ell.|\ell_{i,i-3}|.\n}{d(\ell_{i,i-3})\spab {i-2}.|\ell_{i,i-3}|.\n}.
\end{eqnarray}
Since we have made the choice $\n=i-1$, we find 
\begin{eqnarray}
\mc{G}^i_{i-2}(P_{i,i-3})=\frac{\spab\ell.|\ell_{i,i-3}|.{i-1}}{d(\ell_{i,i-3})\spb \ell.{(i-1)}},
\end{eqnarray}
which is free of spurious singularities. Similarly,  
\begin{eqnarray}
\mc{G}^i_j(P_{i,j})-\mc{G}^i_{j}(P_{i,j-1})&=&\frac{\spa j.\ell}{\spab j.|\ell_{P_{a,b}}|.\n}\frac{d(\ell_{i,j-1})\spab\ell.|\ell_{i,j}|.\n-d(\ell_{i,j})\spab\ell.|\ell_{i,j-1}|.\n}{d(\ell_{i,j})d(\ell_{i,j-1})}\nonumber\\
&=&\frac{\spa j.\ell\spab\ell.|\ell_{i,j}|.j}{d(\ell_{i,j})d(\ell_{i,j-1})},
\end{eqnarray}
so that~\eqref{eq:sum} becomes, 
\begin{eqnarray}
\mc{D}^{(a)}_{i,i-1}&=&-A^{(0)}_n\int \frac{d^4\ell}{(2\pi)^4}\delta(\ell^2)\frac{\spa \im.i}{\spa \ell.i\spa\im.\ell}\bigg(\sum_{j=3}^{i-3}\frac{\spa j.\ell\spab\ell.|\ell_{i,j}|.j}{d(\ell_{i,j})d(\ell_{i,j-1})}\nonumber\\&&-\frac{\spab\ell.|\ell_{i,i-3}|.{i-1}}{d(\ell_{i,i-3})\spb \ell.{(i-1)}}+\mc{G}^i_2(P_{i,2})\bigg).
\end{eqnarray}

In a similar fashion, we find that the contribution from 
Fig.~\ref{fig:N4}$(b)$ has the form
\begin{eqnarray}
\mc{D}^{(b)}_{i,i-1}&=&-A^{(0)}_n\int \frac{d^4\ell}{(2\pi)^4}\delta(\ell^2)\frac{\spa \im.i}{\spa \ell.i\spa\im.\ell}\bigg(\sum_{j=i+1}^{n}\frac{\spa j.\ell\spab\ell.|\ell_{i,j}|.j}{d(\ell_{i,j})d(\ell_{i,j-1})}\nonumber\\&&+\frac{\spab \ell.|\ell_i|.{i-1}}{d(\ell_i)\spb\ell.{(i-1)}}-\mc{G}^i_1(P_{i,n})\bigg).
\end{eqnarray}

Fig.~\ref{fig:N4}$(c)$ has a fixed ordering of gluons and therefore there is no summation over external particles.  There is however a freedom to sum over the particle content of the $\mc{N}=4$ SYM multiplet. We define, 
\begin{eqnarray}
\mc{D}^{(c)}=-\int \frac{d^4\ell}{(2\pi)^4}\delta(\ell^2)\frac{N_i}{d(\ell_{1,i})\spa \ell.i\spab 1.|\ell_{i,1}.\n\spab \ell.|\ell_{i,1}|.\n^2\spab 2.|\ell_{i,1}|.\n\spa {(i-1)}.\ell}\frac{1}{D},
\end{eqnarray}
with,
\begin{eqnarray}
D=\spa i.\ip\dots\spa n.1\spa2.3\dots\spa {i-2}.\im.
\end{eqnarray} 
Here $N_i$ is dependent on the species of loop particle,
\begin{eqnarray}
N_g=\alpha^4+\beta^4, \qquad
N_f=-\alpha^3\beta-\beta^3\alpha, \qquad N_s=2\alpha^2\beta^2,
\end{eqnarray}
where
\begin{eqnarray}
\alpha=\spa1.\ell\spab 2.|\ell_{1,i}|.\n \qquad \mathrm{and} \qquad \beta=\spa2.\ell \spab 1.|\ell_{1,i}|.\n.
\end{eqnarray}
The Schouten identity relates $\alpha$ and $\beta$ to each other, $$\alpha-\beta=\spa1.2\spab\ell.|\ell_{1,i}|.\n\equiv \gamma,$$ 
so that the contribution from the $\mc{N}=4$ multiplet is given by 
\begin{eqnarray}
N_g+4N_f+N_s=\gamma^4.
\end{eqnarray}
Therefore,  the sum of contributions from Fig.~\ref{fig:N4}$(c)$ simplifies to, 
\begin{eqnarray}
\mc{D}^{(c)}&=&-\int \frac{d^4\ell}{(2\pi)^4}\delta(\ell^2)\frac{\spa1.2^4\spab\ell.|\ell_{1,i}|.\n^2}{d(\ell_{1,i})\spa \ell.i\spab 1.|\ell_{i,1}.\n\spab 2.|\ell_{i,1}|.\n\spa {(i-1)}.\ell}\frac{1}{D}\nonumber \\
&=&-A^{(0)}_n\int \frac{d^4\ell}{(2\pi)^4}\delta(\ell^2)\frac{\spa\im.i}{\spa \ell.i\spa {(i-1)}.\ell}\bigg(\mc{G}_1^i(P_{1,i})-\mc{G}_2^i(P_{1,i})\bigg).
\end{eqnarray}
When we combine $(c)$ with $(a)$ and $(b)$ we find the combinations
 $(\mc{G}_1^i(P_{i,n}) - \mc{G}_1^i(P_{i,1}))$ and 
$(\mc{G}_2^i(P_{i,2})-\mc{G}_2^i(P_{i,1}))$ which 
ensure that all spurious singularities are cancelled so that the 
total cut amplitude can be written in a form free of spurious terms, 
\begin{eqnarray}
\mc{D}^{(a)+(b)+(c)}=-A^{(0)}_n\int \frac{d^4\ell}{(2\pi)^4}\delta(\ell^2)\bigg(\sum_{j=i+1}^{i-3}\bigg(\frac{\trm(j,i,\ell,P_{i,j})}{d(\ell_i)d(\ell_{i,j})d(\ell_{i,j-1})}\nonumber\\
+\frac{\trm(j,\im,\ell,P_{i,j})}{d(\ell_{i,i-2})d(\ell_{i,j})d(\ell_{i,j-1})}\bigg)
+\frac{s_{i,i-1}}{d(\ell_i)d(\ell_{i,i-2})}-\frac{\trm(\im,i,\ell,\ell_{i,i-3})}{d(\ell_{i,i-3})d(\ell_i)d(\ell_{i,i-2})}\bigg).
\label{eq:gen}
\end{eqnarray}
The traces can be simplified using the following identity
\begin{eqnarray}
\label{eq:tr}
\trm(j,i,\ell,P_{i,j})&=&2\bigg((\ell.P_{i,j})(i.j)-(\ell.j)(i.P_{i,j})+(\ell.i)(j.P_{i,j})\bigg)\nonumber\\
&=& d(\ell_{i,j})(i.P_{i,j-1})-d(\ell_{i,j-1})(i.P_{i,j})-d(\ell_i)(j.P_{i,j})
+N(P_{i,j},i,j)\nonumber \\
\end{eqnarray}
since the term containing the $\epsilon$ tensor will integrate to zero. 
In eq.~\eqref{eq:tr}, we have introduced the shorthand 
\begin{eqnarray}
N(P,i,j)=P^2(i.j)-2(P.i)(P.j).
\end{eqnarray}

When \eqref{eq:tr} is inserted into \eqref{eq:gen}, there are multiple cancellations of triangle contributions, such that only box contributions remain  
\begin{eqnarray}
\mc{D}^{(a)+(b)+(c)}=-A^{(0)}_n\int \frac{d^4\ell}{(2\pi)^4}\delta(\ell^2)\bigg(\sum_{j=i+2}^{i-3}\bigg(\frac{N(P_{i,j},i,j)}{d(\ell_i)d(\ell_{i,j})d(\ell_{i,j-1})}\nonumber\\
+\frac{N(P_{i,j},\im,j)}{d(\ell_{i,i-2})d(\ell_{i,j})d(\ell_{i,j-1})}\bigg)-\frac{1}{2}\frac{s_{i,i-1}s_{i,i+1}}{d(\ell_{i,i-2})d(\ell_{i,i+1})d(\ell_{i})}
-\frac{1}{2}\frac{s_{i-2,i-1}s_{i,i-1}}{d(\ell_{i,i-3})d(\ell_i)d(\ell_{i,i-2})}\bigg).
\end{eqnarray}
Returning to the Feynman integral by re-instating the cut propagator and the usual normalisation of $i$ we find that the contribution to the amplitude from this cut is 
\begin{eqnarray}
\lefteqn{
A^{(1),\mc{N}=4}(1^-,2^-,\dots,n^+)_{i,i-1~cut}=c_{\Gamma}A^{(0)}_n\bigg\{}\nonumber \\
&&
\sum_{j=i+3}^{i-3}\Ftme(s_{i,j},s_{i+1,j-1};s_{i,j-1},s_{i+1,j}) 
+\sum_{j=i+2}^{i-4}\Ftme(s_{i-1,j},s_{i,j-1};s_{i-1,j-1},s_{i,j})\nonumber\\
&&+\Fom(s_{i,i-2};s_{i,i-1},s_{i-1,i-2})
+\Fom(s_{i+1,i-1};s_{i,i-1},s_{i+1,i})\nonumber\\
&&+\Fom(s_{i,i+2};s_{i,i+1},s_{i+1,i+2})
+\Fom(s_{i-1,i-3};s_{i-1,i-2},s_{i-2,i-3})\bigg\}.
\end{eqnarray}
As expected, we find four one-mass boxes and sums over two-mass easy boxes
in which $p_i$ and $p_{i-1}$ are emitted from different vertices. 

We now know the form of any cut involving two positive helicity gluons.
The mixed helicity, $C_{2,1}$ and $C_{1,n}$, cuts are straightforward,
and it is simple to seethat in these cases the tree factorises in precisely the same manner as before, and that the integrand has exactly the same structure as a $C_{i,i-1}$ cut. 
Therefore to obtain the complete amplitude we need only to sum over all the allowed boxes taking care not to double count any terms i.e.
\begin{eqnarray}
A^{(1),\mc{N}=4}(1^-,2^-,\dots,n^+)=c_{\Gamma}A^{(0)}_n\bigg\{\sum_{i=1}^n\Fom(s_{i,i+2};s_{i,i+1},s_{i+1,i+2})\nonumber\\
+\frac{1}{2}\sum_{i=1}^{n}\sum_{j=i+3}^{n+i-3}
\Ftme(s_{i,j},s_{i+1,j-1};s_{i+1,j},s_{i,j-1})\bigg\} 
\label{eq:Gqcdmhv}
\end{eqnarray}
which agrees with the known result of eq.~\eqref{eq:neq4known}.

\section{Conclusions} 

The calculation of one-loop scattering amplitudes in massless gauge theories can
be simplified by using generalised unitarity constraints to directly extract
coefficients of the one-loop master integrals.  In this paper, we have studied a
further implementation of generalised  unitarity based on putting a single loop
propagator on-shell. Cutting an one-loop $n$-particle amplitude in this way
requires $(n+2)$-particle tree amplitudes as input which are straightforward to
evaluate. If four-dimensional tree amplitudes are used, only the
cut-constructible pieces of the amplitude can be reconstructed. However, if the
cut is performed in  $D=4-2\epsilon$ dimensions then it is possible to fully
determine the amplitude.  As in many of the existing multiple-cut unitarity
methods, once the cut-contribution is in its most simple form, we can reverse
the procedure to find the contribution to the full amplitude from this
particular cut.   For an $n$-point amplitude, one needs fewer than $n$ cuts to
determine the entire amplitude. Furthermore, since the integrand of the cut-loop
integral is itself an $(n+2)$ tree level amplitude,  gauge invariance is assured
for each cut. In this way each single cut contribution is independent of any
other cut, and as such we can use additional cuts as checks on the coefficients
already determined. 

Another advantage of our approach is that one can apply it in either four or
$D$ dimensions, and it can be successfully applied to amplitudes that are fully
cut-constructible, amplitudes that are entirely rational, or amplitudes that
have both contributions. We have tested the validity of our method by
re-deriving the entirely rational one-loop all-plus amplitudes for four and five
gluons,  $A_4^{(1)}(+,+,+,+)$ and $A_5^{(1)}(+,+,+,+,+)$, as well as the
mostly-plus  four-gluon amplitude $A_4^{(1)}(-,+,+,+)$.  The key ingredient is
to relate the $D$-dimensional cut for massless loop particles to a
four-dimensional cut for massive particles.  For this task we used the CSW
prescription with massive scalars recently derived by Boels and Schwinn
\cite{Boels:2007pj,Boels:2008ef}. When constructing the integrand we frequently
encountered terms of the form $\spab i.|\ell_i|.\n$ which do not correspond to
any physical propagator (they are spurious terms). Before evaluating any
integrals it was essential to write the integrand in a form free of these terms.
We found that in general there was a clear diagrammatic way of combining diagrams
to remove these terms. After this reduction had occurred we found that the
remaining terms usually had a simple structure allowing quick reconstruction of
the amplitude. Nevertheless, although the CSW prescription provides a clear
diagrammatic prescription of the integrand, it is not an essential part of the
method. 

We have also rederived expressions for the scalar loop contribution to the
four-gluon MHV amplitude, $A_4^{(1,\mc{N}=0)}(-,-,+,+)$ which has both
cut-constructible and rational contributions, and the fully cut-constructible
$n$-gluon MHV amplitude in $\mc{N}=4$ SYM, $A_4^{(1,\mc{N}=4
)}(-,-,+,\ldots,+)$.

A final advantage of the single cut unitarity approach, is that it is the only
unitarity-based technique that can detect the coefficients of tadpole graphs.
Although in this work we have concentrated on solely on massless theories, for
which these terms are absent,  the prospect of applying the single cut approach
to a massive theory may be worthy of further study.

\acknowledgments 

We are grateful to Pierpaolo Mastrolia for many useful conversations, and collaboration at an early stage in the project. 
CW acknowledges
the award of an STFC studentship. EWNG gratefully acknowledges the support of the Wolfson Foundation and the Royal Society

\appendix

\section{Notation and Scalar integrals}

Throughout this paper we use the following notation
\begin{eqnarray}
P_{i,j}&=&p_i+p_{i+1}+\dots+p_{j-1}+p_j,\\
s_{i,j}&=&P^2_{i,j},\\
s_{ij}&=&2(p_i\cdot p_j).
\end{eqnarray}
For the four point amplitudes $-+++$ and $--++$ we also use the notation, 
\begin{eqnarray}
s&=&(p_1+p_2)^2=(p_3+p_4)^2,\\
t&=&(p_2+p_3)^2=(p_1+p_4)^2,\\
u&=&(p_2+p_4)^2=(p_1+p_3)^2.
\end{eqnarray}
During the construction of $D$-dimensional amplitudes we encounter the following functions, 
\begin{eqnarray}
I_n[\mu^2]=J_n=(-\epsilon)I_n^{6-2\epsilon}  \qquad \mathrm{and} \qquad I_4[\mu^4]=K_4=(-\epsilon)(1-\e)I_4^{8-2\epsilon}.
\end{eqnarray}
These integrals have the following form,
\begin{eqnarray}
I^{6-2\e}_2(P^2)=-\frac{r_{\Gamma}}{2\e(1-2\e)(3-2\e)}(-P^2)^{1-\e},\\
I^{6-2\e}_3(P^2)=-\frac{r_{\Gamma}}{2\e(1-\e)(1-2\e)}(-P^2)^{1-\e}.
\end{eqnarray}
Here $r_{\Gamma}=c_{\Gamma}/(4\pi)^{2-\epsilon}$. The boxes we find have the following $\e\rightarrow 0$ expansion,
\begin{eqnarray}
J_4\rightarrow0+\mc{O}(\e) \qquad \mathrm{and} \qquad K_4\rightarrow-\frac{1}{6}+\mc{O}(\e).
\end{eqnarray}
Finally, the ten dimensional pentagon we encounter has the following limit as $\e\rightarrow 0$
\begin{eqnarray}
\e(1-\e)I_5^{10-2\e}\rightarrow \frac{1}{24} +\mc{O}(\e).
\end{eqnarray}
The two-mass easy and one mass box functions which arise in the calculation of the $\mc{N}=4$ MHV amplitude have the following form,  
\begin{align}
	F^{1m}_4(P^2;s,t) = -\frac{1}{\e^2}
	\bigg[
	&\left (\frac{\mu^2}{-s}\right )^{\e}\FF{1,-\e;1-\e;-\frac{u}{t}} \nonumber\\
	+&\left (\frac{\mu^2}{-t}\right )^{\e}\FF{1,-\e;1-\e;-\frac{u}{s}} \nonumber\\
	-&\left (\frac{\mu^2}{-P^2}\right )^{\e}\FF{1,-\e;1-\e;-\frac{uP^2}{st}} 
	\bigg],
	\label{eq:f1m}\\
	F^{2me}_4(P^2,Q^2;s,t)  = -\frac{1}{\e^2}
	\bigg[
	&\left (\frac{\mu^2}{-s}\right )^{\e}\FF{1,-\e;1-\e;
	\frac{
	us
	}{
	P^2Q^2-st
	}} \nonumber\\
	+&\left (\frac{\mu^2}{-t}\right )^{\e}\FF{1,-\e;1-\e;
	\frac{
	ut
	}{
	P^2Q^2-st
	}} \nonumber\\
	-&\left (\frac{\mu^2}{-P^2}\right )^{\e}\FF{1,-\e;1-\e;
	\frac{
	uP^2
	}{
	P^2Q^2-st
	}} \nonumber\\ 
	-&\left (\frac{\mu^2}{-Q^2}\right )^{\e}\FF{1,-\e;1-\e;
	\frac{
	uQ^2
	}{
	P^2Q^2-st
	}}
	\bigg].
	\label{eq:f2me}
\end{align}

\bibliographystyle{JHEP-2}
\providecommand{\href}[2]{#2}\begingroup\raggedright\endgroup

\end{document}